\documentclass[twocolumn,twocolappendix]{aa}

\usepackage[varg]{txfonts}
\usepackage{graphicx}
\usepackage{natbib}
\usepackage{xspace}
\usepackage[colorlinks,citecolor=blue,linkcolor=blue,urlcolor=blue]{hyperref} 

\newcommand{\Msun}{\ensuremath{\, { M}_{\odot}}}
\newcommand{\msun}{\ensuremath{\, { M}_{\odot}}}
\newcommand{\Lsun}{\ensuremath{\, { L}_{\odot}}}
\newcommand{\lsun}{\ensuremath{\, { L}_{\odot}}}
\newcommand{\logL}{\ensuremath{\log_{10}(L/  L_{\odot})}}

\newcommand{\Myr}{\ensuremath{\,\mathrm{Myr}}\xspace}
\newcommand{\Myrs}{\ensuremath{\,\mathrm{Myr}}\xspace}

\newcommand{\posydon}{\texttt{POSYDON}\xspace}
\def\mesa{\texttt{MESA}}
\newcommand{\fig}[1]{Fig.~\ref{#1}}

\newcommand{\yangt}{Y23\xspace}

\def\dejager{$\dot{M}_{\rm J88}$\xspace}
\def\yang{$\dot{M}_{\rm Y23}$\xspace}
\def\beasor{$\dot{M}_{\rm B20}$\xspace}
\def\kee{$\dot{M}_{\rm K21}$\xspace}
\def\fecc{FeCC\xspace}

\def\ecsne{ECSNe\xspace}

\def\plotfolder{plotsMarch2024} 

\begin{document}

\title{The effect of mass loss in models of red supergiants in the Small Magellanic Cloud}

  \author{E.~Zapartas\inst{\ref{forth}, \ref{noa}},
  S.~de Wit\inst{\ref{noa}, \ref{nkua}}, 
  K.~Antoniadis\inst{\ref{noa}, \ref{nkua}},
  G.~Mu\~{n}oz-Sanchez\inst{\ref{noa}, \ref{nkua}}, 
   D.~Souropanis\inst{\ref{forth}},
  A.Z.~Bonanos\inst{\ref{noa}}, 
G.\,Maravelias\inst{\ref{noa},\ref{forth}}, 
K.\,Kovlakas\inst{\ref{ice},\ref{ieec}},  
M.\,U.\,Kruckow\inst{\ref{geneva},\ref{gwsc}}, 
T.~Fragos\inst{\ref{geneva},\ref{gwsc}}, 
J.~J.~Andrews\inst{\ref{florida}}, 
S.\,S.\,Bavera\inst{\ref{geneva},\ref{gwsc}}, 
M.\,Briel\inst{\ref{geneva},\ref{gwsc}}, 
S.\,Gossage\inst{\ref{ciera}}, 
E.\,Kasdagli\inst{\ref{florida}},
K.~A.~Rocha\inst{\ref{nw},\ref{ciera}}, 
M.\,Sun\inst{\ref{ciera}}, 
P.\,M.\,Srivastava\inst{\ref{ciera},\ref{electrical_nw}}, 
Z.\,Xing\inst{\ref{geneva},\ref{gwsc},\ref{ciera}} 
}




\institute{Institute of Astrophysics FORTH, 71110 Heraklion, Greece\label{forth}
\and
IAASARS, National Observatory of Athens, Vas. Pavlou and I. Metaxa, Penteli, 15236, Greece\label{noa}
\and
Department of Physics, National and Kapodistrian University of Athens, Panepistimiopolis, Zografos, 15784, Greece\label{nkua}
\and        
Département d'Astronomie, Université de Genève, Chemin Pegasi 51, CH-1290 Versoix, Switzerland\label{geneva}
\and        
Gravitational Wave Science Center (GWSC), Université de Genève, CH1211 Geneva, Switzerland\label{gwsc}
\and
Department of Physics, University of Florida, 2001 Museum Rd, Gainesville, FL 32611, USA\label{florida}
\and
Center for Interdisciplinary Exploration and Research in Astrophysics (CIERA), 1800 Sherman, Evanston, IL 60201, USA\label{ciera}
\and
Institute of Space Sciences (ICE, CSIC), Campus UAB, Carrer de Magrans, 08193 Barcelona, Spain\label{ice}
\and
Institut d'Estudis Espacials de Catalunya (IEEC),  Edifici RDIT, Campus UPC, 08860 Castelldefels, Barcelona, Spain\label{ieec}
\and
Department of Physics \& Astronomy, Northwestern University, 2145 Sheridan Road, Evanston, IL 60208, USA\label{nw}
\and
Electrical and Computer Engineering, Northwestern University, 2145 Sheridan Road, Evanston, IL 60208, USA\label{electrical_nw}
}

\authorrunning{Zapartas et al.}

\abstract{
The rate and mechanism of mass loss of red supergiants  (RSGs) remain poorly understood, especially at low metallicities.  
Motivated by the new empirical prescription by Yang et al. 2023, based on the largest and most complete sample in the Small Magellanic Cloud, we investigate the impact of different popular and recent  RSG mass-loss prescriptions that span a range of RSG mass-loss rates on the evolution and observable properties of single massive stars. 
%
Our results show that higher mass-loss rates result in earlier envelope stripping and shorter RSG lifetimes, particularly for the more luminous stars, leading to a  steeper luminosity function and predicting hotter final positions for the SN progenitors. None of the considered mass-loss prescriptions is fully consistent with all observational constraints, highlighting ongoing uncertainties in deriving and modeling RSGs mass loss.  
The mass-loss rates suggested by Kee et al. predict rapid envelope stripping, inconsistent with the observed population of luminous RSGs and SN progenitor detections, while the models implementing the commonly used de Jager et al.\ and the recent Beasor et al.\ prescriptions overestimate the number of luminous RSGs. While the increased mass-loss rates for luminous RSGs predicted by Yang et al. lead to better agreement with the observed RSG luminosity function, naturally reproducing the updated Humphreys-Davidson limit, they also produce luminous yellow supergiant progenitors not detected in nearby supernovae. 
%
We also estimate that binary interactions tend to slightly increase the formation of luminous RSGs due to mass accretion or merging. 
%
Our study examines the impact of RSG mass loss during the late stages of massive stars, highlighting the significance of using comprehensive observational data, exploring the uncertainties involved, and considering the effects of binary-induced or episodic mass loss. 
}


\date{}

\titlerunning{The effect of mass loss in RSGs of the SMC}

\maketitle



\section{Introduction}\label{sec:intro}


Massive stars with initial masses in the range of  $\sim 8-30 \  \Msun$ become red supergiants (RSGs) during their late evolutionary phases, following their departure from the main sequence \citep[e.g.,][and references therein]{Levesque2017}. 
RSGs reside on the cool, luminous region of the Hertzsprung-Russell diagram (HRD). 
 For stars within this mass range, mass loss during the main-sequence phase, where they spend most of their lifetime, is generally minimal \citep[although important for certain evolutionary outcomes as discussed in][]{Renzo+2017}. It is during the RSG phase that these stars experience substantial mass loss, profoundly influencing their subsequent evolution and ultimate fate \citep[e.g.,][]{Meynet+Maeder2003, Meynet+2015, Beasor+2020}.

Eventually, most RSGs are expected to end their lives in Type II core-collapse supernovae \citep[SNe; e.g.,][]{Heger+2003, Langer2012}. This has been confirmed observationally in pre-SN progenitor detections of nearby events \citep[e.g.,][]{Smartt+2009,Smartt+2015}. However, some may directly collapse into black holes with or without a faint transient event \citep{OConnor+2011}, a phenomenon with possible observed candidates \citep{Adams+2017, Beasor+2024}. This possibility has been proposed as the explanation of the ``Red-supergiant problem", i.e. lack of detections of luminous, high-mass RSG progenitors \citep[][although see a discussion about the low statistical significance of the issue in \citealt{Davies+2018}]{Smartt+2009}. 
An alternative solution consists of strong winds that partially strip the H-rich envelope of luminous RSGs, which become hotter and bluer and avoid exploding as Type II SNe \citep[e.g.,][]{Georgy+2013,Meynet+2015}. Consequently, mass loss affects the evolution of a massive star and its position in the HRD, its final mass, the remnant formed during core collapse, and the observable characteristics of its eventual supernova explosion or black hole formation. 

Although recent studies contribute to constraining the mass-loss rate empirically \citep[e.g.][]{Beasor+2020, Davies+2021, Decin+2024, Yang+2023, Antoniadis+2024}, or even theoretically \citep{Kee+2021, Vink+2023,Fuller+2024}, 
the driving mechanism and the rate of mass loss during that phase remains uncertain \citep[e.g.,][]{Yoon+Cantiello2010,Smith2014,Arroyo-Torres+2015}, with indications for episodic mass loss \citep[e.g.,][]{Decin+2006, Bruch2021, Dupree+2022, Humphreys2022, Cheng+2024} also complicating the picture. 
To include the significant effect of mass loss during the RSG phase, stellar evolution models need to select from a wide range of predominantly empirical mass-loss rate prescriptions \citep{Reimers1975,de-Jager+1988, Nieuwenhuijzen+1990, van-Loon+2005,Goldman+2017,Beasor+2020, Kee+2021, Antoniadis+2024} that vary by orders of magnitude. 
\citet{van-Loon+2005,Goldman+2017,Yang+2023} predict higher mass-loss rates, mainly due to the assumption of radiatively-driven wind being the mechanism of mass loss, in comparison with the weaker winds found when steady-state density distribution is considered   \citep{Beasor+2020,Antoniadis+2024}. Most of these studies are based on the properties of the dust shell formed around the RSG. Different ranges of the assumed grain sizes can also affect the inferred mass-loss rate \citep{Antoniadis+2024}. Furthermore, a uniform average gas-to-dust ratio is usually assumed, which can vary significantly within some galaxies, such as the SMC \citep{Clark+2023}. Recent alternative methods using gas and molecular diagnostics \citep{Decin+2024,Gonzalez-Tora+2024} inferred also a wide range of mass-loss rates. 
%
Incorporating accurate mass loss prescriptions into stellar evolution models is essential for predicting the late stages of massive stars, their fate, and understanding their roles in broader astrophysical processes. 


Especially in low metallicity regimes, despite efforts to constrain RSG properties \citep[e.g.,][]{Levesque+2006,Davies+2015,Patrick+2017,Britavskiy+2019a,Gonzalez-Torra+2021,de-Wit+2023,Bonanos+2024},
the mass-loss rate at this phase has been less constrained (with the exception of the study by \citealt{van-Loon+2005}), but there has been significant progress in the last decade  \citep{Yang+2023, Antoniadis+2024}. 
\citet[][hereafter \yangt]{Yang+2023} analyzed a comprehensive RSG sample in the Small Magellanic Cloud (SMC) and discovered a notable "kink", i.e. a change in slope in the mass-loss rate as a function of luminosity where the winds significantly increase for RSGs with $\log(L/L_\odot) \gtrsim 4.6$. This feature suggests that more luminous RSGs experience a higher mass-loss rate, potentially leading to earlier envelope stripping and shorter RSG lifetimes. The same feature was also found by \citet{Antoniadis+2024} for the RSGs in the Large Magellanic Cloud, although at slightly lower luminosity value.  Motivated by the new, empirical mass-loss rate relation presented by \yangt, we proceed to implement this and other mass loss prescriptions in the literature, in stellar evolution models of SMC metallicity to compare theoretical stellar populations with observations, investigating the impact of stellar mass loss on the life and the eventual death of RSGs.

The paper is structured as follows: In Sect.~\ref{sec:method} we present the different RSG mass-loss rate prescriptions implemented in our stellar models, as well as the observational sample used for comparison. In Sect.~\ref{sec:results} we show the effects of the RSG mass-loss rate on RSG evolution, their luminosity function and their final fate. We estimate the possible effect of binaries, discuss the yellow-to-red supergiants ratio, the tension with observable constraints within the caveats of our study, as well as the feedback from RSGs into its environments in Sect.~\ref{sec:discussion}. We provide some concluding thoughts in Sect.~\ref{sec:conclusions}


\section{Method}\label{sec:method}

\subsection{Stellar tracks and simulations of a RSG population, with \posydon} \label{sec:simulation_SMC}

 To investigate the effect of different RSG mass-loss rates in the evolution and final outcome of massive stars, we implement the \posydon\footnote{\url{https://posydon.org/}} 
framework \citep{Fragos+2023,Andrews+inprep}, which is based on grids of detailed single- and binary-star models computed with the MESA 1D stellar evolution code \citep[][version 11701]{Paxton+2011,Paxton+2013, Paxton+2015, Paxton+2018, Paxton+2019, Paxton+2021} from zero-age main sequence up to carbon depletion in the core. 
Our results are based on single-star model grids, spanning initial masses from 
6 to 40 \Msun, with a logarithmic spacing of $\sim 0.0093$, ensuring that we include all possible single massive star tracks that may pass through the RSG phase during their lifetime \citep[e.g.,][]{Heger+2003, Ekstrom+2012}. 
The default physical assumptions and numerical specifications follow those of \posydon~v2 stellar models \citet{Andrews+inprep}, apart from the three extra different mass-loss rate prescriptions during the RSG phase, as we describe in Sect.~\ref{sec:Mdot_prescriptions}. 
Although there is a scatter of metallicities across the SMC \citep{Davies+2015, Choudhury+2018}, we assume an average metallicity of $Z=0.2\, Z_{\odot} = 0.00284\, ({\rm [Fe/H]} \sim -0.7)$, close to the values from previous stellar modeling of the SMC \citep{Brott+2011, Georgy+2013}. The solar metallicity calibration is based on \citet[][$Z_{\odot}=0.0142$]{Asplund+2009}.

A significant fraction of the RSGs in our sample are expected to be products of prior binary evolution \citep{Sana+2012,Moe+2017}, involved in scenarios of merging, mass accretion, or even partial stripping \citep[e.g.,][]{deMink+2014,Justham+2014,Eldridge+2019}. 
However, for simplicity, as well as in order to obtain a better handle on the differences among the various mass-loss rate assumptions explored, in this paper we focus on the single-star channel as a first step. We estimate the impact of a possible binary history of RSGs in Sect.~\ref{sec:binarity}.


%

We aim to prevent contamination of the RSG sample with asymptotic giant branch (AGB) stars, however, distinguishing between these types of stars photometrically is challenging. Previously, \citet{Yang+2023} implemented various observational cuts and criteria to minimize AGB contamination, while \citet{Yang+2024}, who analyzed a large spectroscopic sample, found the lower luminosity limit for RSGs in the Large Magellanic Cloud to be as low as $\log_{10}(L_{\rm min}/ \rm L_{\odot}) \sim 3.5$. 
However, it remains uncertain whether the interior physics and the mass-loss rates of these lower-luminosity and lower-mass stars are analogous to more massive ones, and if they end up exploding as a core-collapse SN. Therefore, in this study, we choose to set a theoretically-motivated limit. AGB and super-AGB stars originate from intermediate-mass stars, that eventually form a carbon-oxygen or (for initially slightly more massive cases) an oxygen-magnesium core that is not massive enough to collapse, forming a white dwarf instead \citep{Poelarends+2008, Doherty+2015}. To be conservative and avoid these scenarios altogether, we only accept tracks that eventually burn up all the heavy elements up to iron in their core, producing an iron-core-collapse (\fecc). We thus also reject progenitors of electron-capture SNe (ECSNe) from oxygen-magnesium cores close to the Chandrasekhar limit, which according to \citet{Podsiadlowski+2004} occur for 
final helium core masses $M_{\rm He,core,fin} < 2.5$~\Msun. This is dependent on the mass loss model we use, but corresponds to a minimum luminosity of $\log_{10}(L_{\rm min}/ \rm L_{\odot}) \sim  4.03$. 

To create a population of RSGs, for each mass-loss assumption simulation, we randomly draw $10^5$ initial masses between $[6-40]$~\Msun, according to a canonical \citep{Kroupa+2001} initial mass function (IMF). 
There is evidence for a slightly more top-heavy IMF in the Magellanic Clouds \citep{Schneider+2018}, 
but it mostly affects stars above $\sim 30\Msun$ so we do not consider it. We linearly interpolate between the stellar tracks of different initial masses \citep[as explained in][]{Dotter2016, Fragos+2023}. 
For each drawn mass we also pick a random time during its evolution between 8 and 70 \Myr, to include all possible stellar lifetimes of single stars that may become RSGs \citep[e.g.,][]{Ekstrom+2012,Zapartas+2017}. As we want to compare with observational findings in the SMC from \yangt (discussed in Sec~\ref{sec:obs_sample}), we weight the drawing according to a star formation history (SFH) of the SMC. We follow \citet{Rubele+2015}, 
which estimates three lookback time bins of slightly different star formation rates:
\begin{align}
\mathrm{SFH_{SMC}}(t) /(M_{\odot} {\rm yr}^{-1}) =
  \begin{cases}
   0.052,&\mathrm{\,\,for\,\,\,} t<14 \\
   0.125,&\mathrm{\,\,for\,\,\,}  t \in [14,40]  \\
   0.06875,&\mathrm{\,\,for\,\,\,} t \geq 40, \\
  \end{cases}
\end{align}
where $t$ is the lookback time in Myr. 
We consider a star to be in its RSG phase when its effective temperature reaches below $T_{\rm eff, max} = 10^{3.66} \simeq 4570$~K \citep{Meynet+2015}.


\subsection {Mass-loss prescriptions}\label{sec:Mdot_prescriptions}
  
\begin{figure*}
\centering
\includegraphics[width=0.9\textwidth]%
{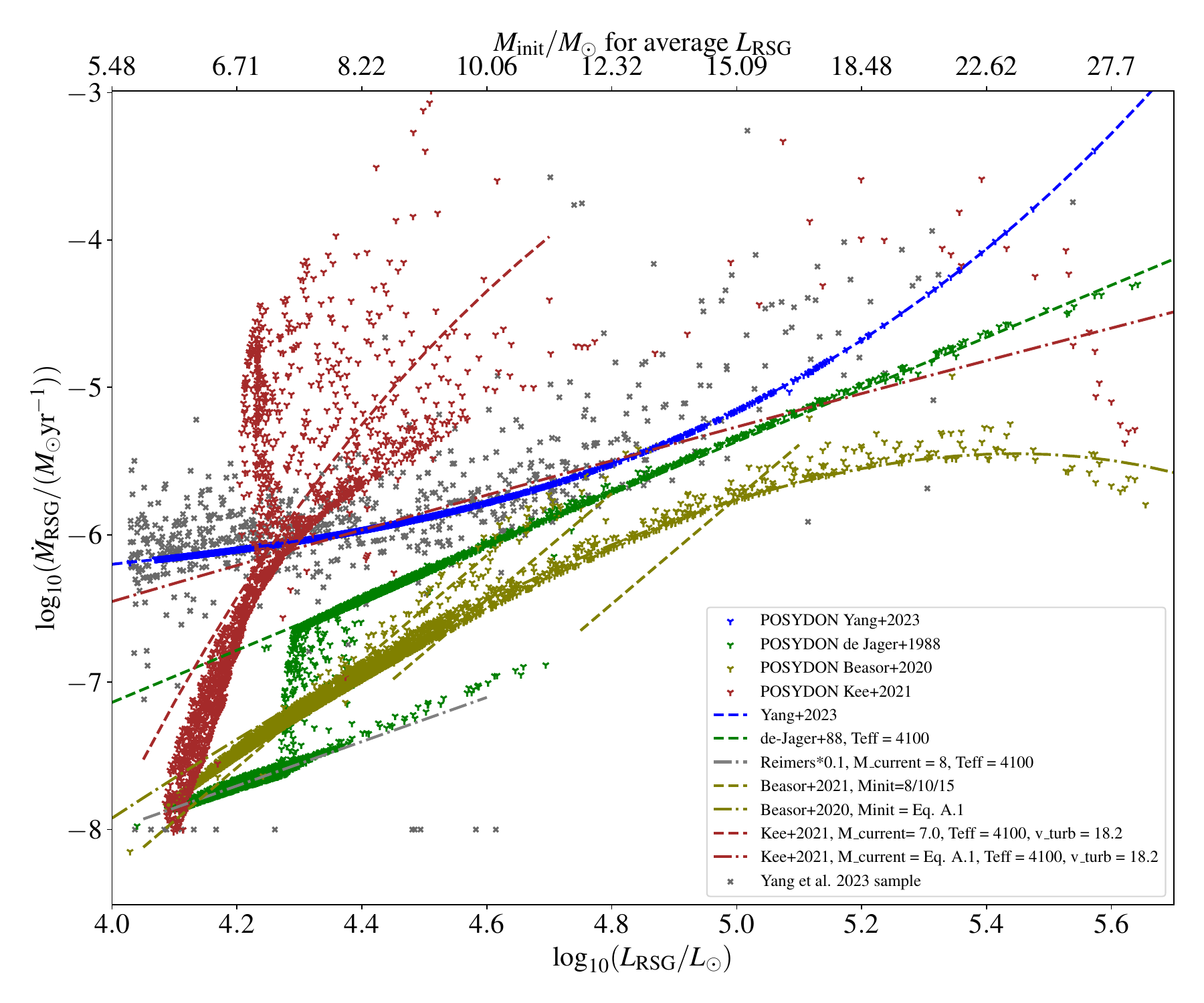}
\caption{Mass-loss rate versus luminosity of the RSG wind prescriptions implemented. Points are random draws of RSGs from \posydon stellar populations, following each prescription (see legend) and grey crosses are the \yangt observational sample. The top $x$-axis corresponds to the initial mass of a star to have $L_{\rm RSG,avg}$ equal to the bottom axis, according to Eq.~\ref{eq:Minit_L}.
} 
\label{fig:Mdot_luminosity}
\end{figure*}

This study focuses on widely used and recent RSG mass-loss prescriptions, described below. These four prescriptions are derived employing different inference techniques, sample selections, and sizes, which in turn span a range of RSG mass-loss rates. 
The four grids of stellar tracks differ solely in the implemented prescription. We show in \fig{fig:Mdot_luminosity} the selected prescriptions as a function of RSG luminosity, with points depicting random draws of RSGs from our \posydon stellar populations, with each color corresponding to a different mass loss prescription.
The concentration of drawn points at low luminosities reflects the weighting by the IMF and the longer RSG lifetimes for lower initial masses. 

\begin{itemize}

\item One selected RSG mass loss prescription is from \citet[][from now on \dejager, green in \fig{fig:Mdot_luminosity}]{de-Jager+1988}, which is one of the most popular prescriptions used in stellar models for this evolutionary phase, and is the default RSG mass-loss prescription used in \posydon. The mass-loss rate of this empirical relation 
is based on the position at the HRD of around a dozen Galactic RSGs, and is calculated 
for effective temperatures below $10^4~{\rm K}$, 


\begin{align}\label{eq:dejager}
\log_{10}(\dot{M}/(M_{\odot} {\rm yr}^{-1})) = &1.769\log_{10}(L/  L_{\odot}) \nonumber \\ 
&-1.676\log_{10}(T_{\rm eff}/{\rm K})  -  8.158.
\end{align}

The reason that some points of \dejager do not fall exactly on top of the depicted lines of the theoretical prescription in \fig{fig:Mdot_luminosity} is that the prescription depends also on $T_{\rm eff}$, apart from luminosity.
%
The drop of some drawn RSGs mass-loss rate luminosities below $\log_{10}(L/L_\odot)\sim 4.3$ for RSG following \dejager is because of the \posydon implementation of \citet{Reimers1975} mass loss prescription (multiplied by a factor of 0.1, grey line) when the total stellar mass drops below 8\Msun, which is considered a more viable option for- low and intermediate-mass stars.

\item We implemented the  
recent RSG mass loss prescription \citet[][from now on \yang]{Yang+2023}, based on the largest sample of RSGs in the SMC, as shown in Eq.~\ref{eq:yang},

\begin{align}\label{eq:yang}
\log_{10}(\dot{M}/(M_{\odot} {\rm yr}^{-1})) &= 20.3\log_{10}(L/ L_{\odot}) - 5.09[\log_{10}(L/  L_{\odot})]^2 \nonumber \\
& - 0.44[\log_{10}(L/  L_{\odot})]^3 - 33.91. 
\end{align}

Since \yang does not have a dependence on the effective temperature, $T_{\rm eff}$, we need to define the region of its validity in the HRD. We thus applied in our models a linear transition from the default \dejager scheme towards \yang prescription between 6000 and 5000 K, testing that our results are not sensitive to the exact transition limits. Due to the dependence only on luminosity, all the drawn RSGs with \yang fall by definition on top of the theoretical prescription in \fig{fig:Mdot_luminosity}.  On the same plot, we also depict with grey crosses the \yangt sample of RSGs, which \yang was based on\footnote{As noted in \citet{Antoniadis+2024}, 
there is a discrepancy between the equation published in \yangt and the correct value which would yield higher mass-loss rates by 0.7 dex, but we followed the published version for consistency.}.

\yang is only a factor of a few higher than \dejager at the mid-range of RSG luminosities,  $4.5 \lesssim \log_{10}(
L/\Lsun) \lesssim 5.1$, but even higher for lower or higher luminosities. 
%
\yang is closer to the mass-loss rate implemented in the standard Geneva models  \citep{Ekstrom+2012} multiplied by ten (``10xMdot''), as investigated in \citet{Meynet+2015}. 

\item We have implemented the mass-loss rate prescription from \citet[][in its corrected form from \citealt{Beasor+2023}; referred to hereafter as  \beasor]{Beasor+2020}, 
which is based on coeval samples of RSGs in Galactic clusters. 
The prescription functions in our study as an example of a low mass-loss rate during the RSG phase:

\begin{align}\label{eq:beasor}
\log_{10}(\dot{M}/(M_{\odot} {\rm yr}^{-1})) &= 4.8\log_{10}(L/ L_{\odot}) \nonumber \\
&- 0.23\log_{10}(M_{\rm init}/M_{\odot})  -   24.6. 
\end{align}

The \beasor prescription depends on the initial mass ($M_{\rm init}$), based on the assumption that the RSGs in their cluster samples have similar initial masses.
For convenience, we use a relation between initial mass, $M_{\rm init}$, and average RSG luminosity, $L_{\rm RSG,avg}$, at various points throughout the study. The exact recalculation of the relation for our models is provided in Appendix~\ref{sec:Minit_LRSG}. Indeed, RSGs with \beasor follow very well the theoretical prescription, if we convolve it with our $M_{\rm init}-L_{\rm RSG,avg}$ best-fit relation of Eq.~\ref{eq:Minit_L}. 

\item Finally, we have also applied the first attempt for a theoretical prescription of RSG mass loss by \citet[][from now on \kee]{Kee+2021}. This work assumes that turbulence is the dominant driving mechanism for mass loss, in line with previous studies \citep[e.g.][]{Josselin+2007}. The \kee prescription is highly sensitive, as expected, to the assumed turbulent velocity on the surface (see their Fig. 3). We use $v_{\rm turb} = 18.2$ km s$^{-1}$, which they find as the best-fit empirical value. 

The model points in \fig{fig:Mdot_luminosity} occupy mostly the low luminosity regime, corresponding to initial masses around $7-8$\Msun. This is because for higher initial masses the stars quickly become stripped and only exist as RSGs for a short time.
The strong dependence on 
the current mass during the RSG phase, 
(see their Fig. 2) creates a runaway effect of increasing mass-loss rate for a RSG which retains its position in the HRD (i.e., maintaining a constant stellar radius). This leads to a significant decrease in mass and, subsequently, surface gravity,  which further intensifies the wind by orders of magnitude. As the star becomes progressively stripped, its hotter inner layers are exposed, ultimately leading to a blueward evolution. 

The theoretical curve of \kee in \fig{fig:Mdot_luminosity} (red dashed-dotted curve), when we convolve it with our best-fit $M_{\rm init}-L_{\rm RSG,avg}$, Eq.~\ref{eq:Minit_L}, becomes similar to the one in Figure 8 of \citet{Kee+2021}. On the other hand,  for fixed mass it becomes much steeper with luminosity (red dashed line in \fig{fig:Mdot_luminosity} for $M= 7 \Msun$). In practice, as we will see in Sec.~\ref{sec:results}, the mass of a RSG with \kee will decrease quickly and its mass-loss rate will keep increasing significantly. 

\end{itemize}


%

We have not implemented a direct dependency on the metallicity in any of the considered RSG mass loss prescriptions.  No clear evidence of such a dependency has been presented to date, despite efforts on theoretical \citep{Kee+2021} and empirical grounds \citep[e.g.,][Antoniadis et al. in prep.]{Goldman+2017}. Therefore we refrain from adding an extra uncertainty into our analysis from a poorly-constrained metallicity dependence of the RSG winds mass-loss rates. Nevertheless, the \yang prescription has been calibrated on SMC data, and thus would not have required a metallicity adjustment.

During a star's blueward evolution, when its $T_{\rm eff}>10^4 {\rm K}$ and its surface hydrogen mass fraction $X_{\rm surf}$ drops below $0.4$, we assume that the Wolf-Rayet line-driven winds kick in \citep[][implemented in \posydon]{Nugis+Lamers2000}, which are stronger than the star's prior winds during its main-sequence evolution \citep{Vink+2000}. 
If a very luminous star enters inside the Humphreys-Davidson limit \citep[HD;][]{Humphreys+Davidson1979}, with $\log_{10}(L_{\rm RSG}/\lsun)\gtrsim 5.78$ for RSGs, following \citet[][although more recent studies suggested a lower luminosity limit; e.g., \citealt{Davies+2018b,McDonald+2022}]{Hurley+2000,Belczynski+2010}, we change the mass-loss rate to a constant value of $10^{-4} \,\Msun \rm{yr}^{-1}$ \citep{Belczynski+2010}, to simulate the regime of high but uncertain ``LBV-like" mass loss. We do not implement the recent prescriptions including extra ``eruptive" mass loss due to local super-Eddington layers at the stellar surface \citep{Cheng+2024}.    
%


\begin{figure*}[t]
\centering
\includegraphics[width=0.5\linewidth]{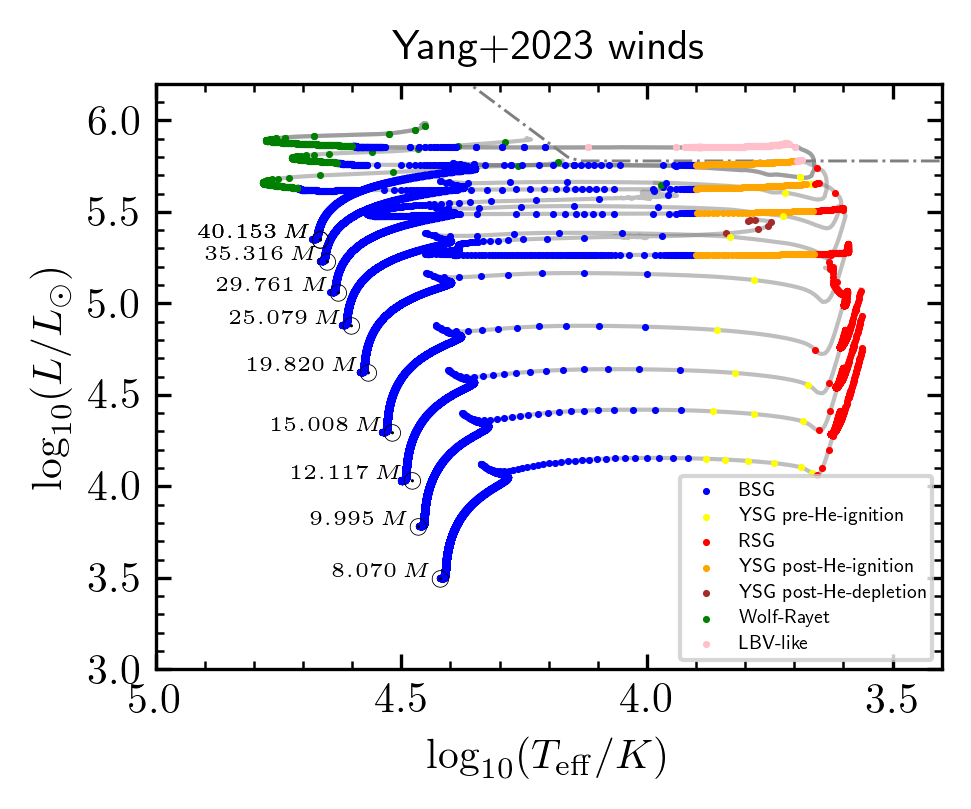}\includegraphics[width=0.5\linewidth]{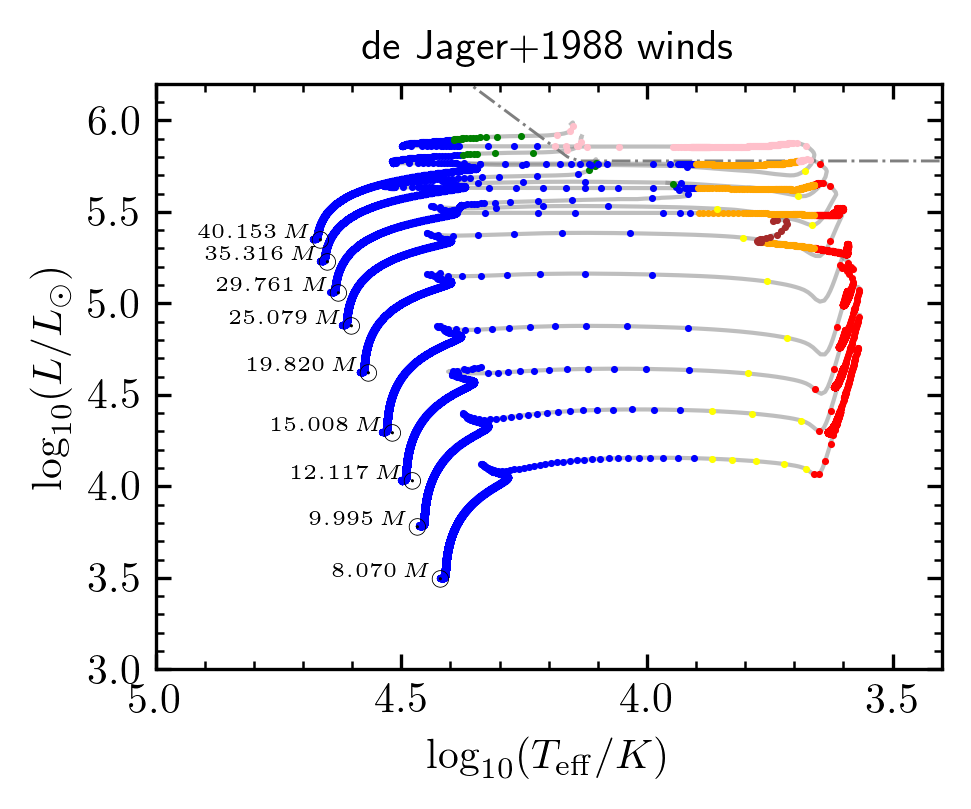}\\
\includegraphics[width=0.5\linewidth]
{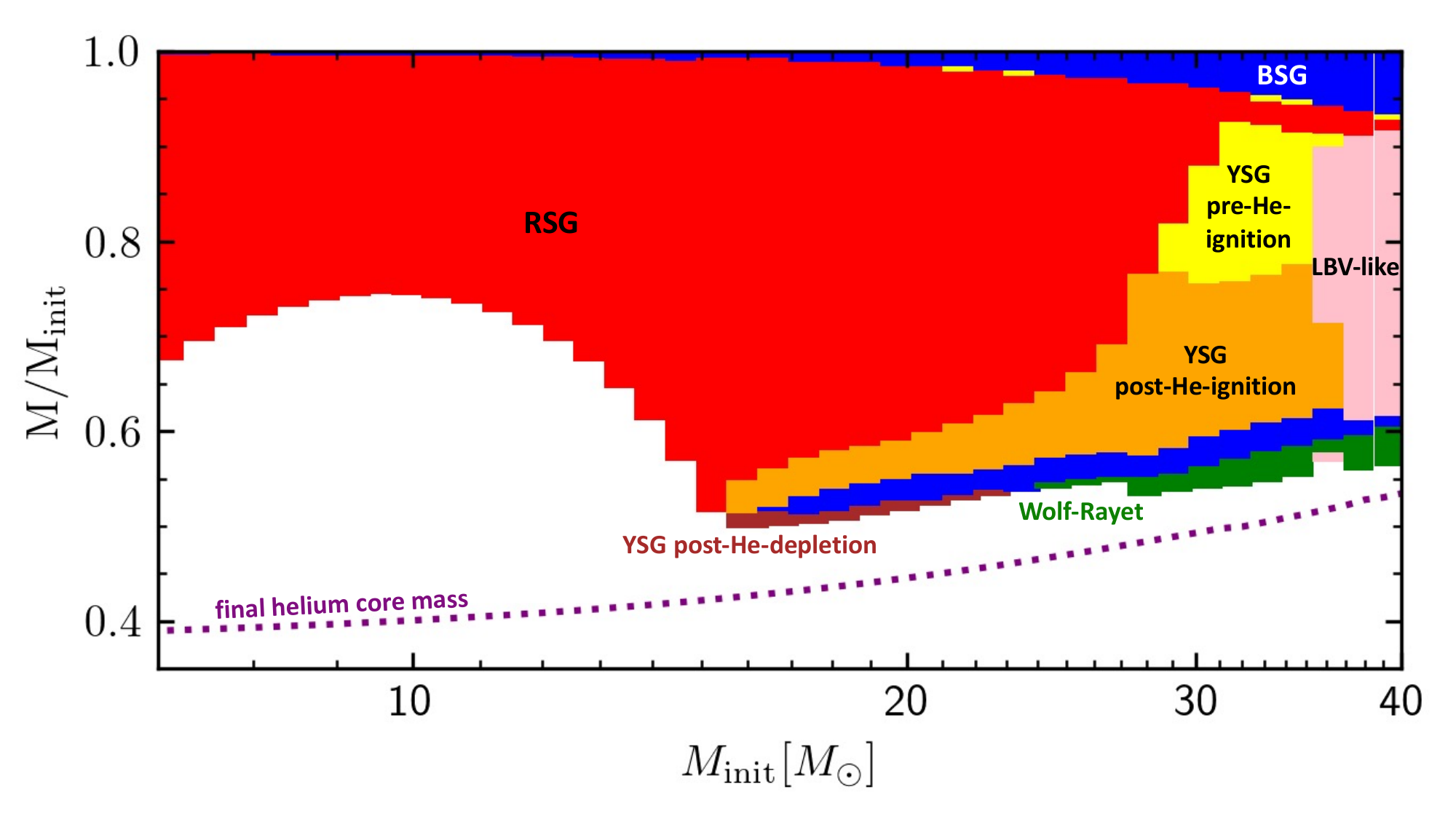}\includegraphics[width=0.5\linewidth]{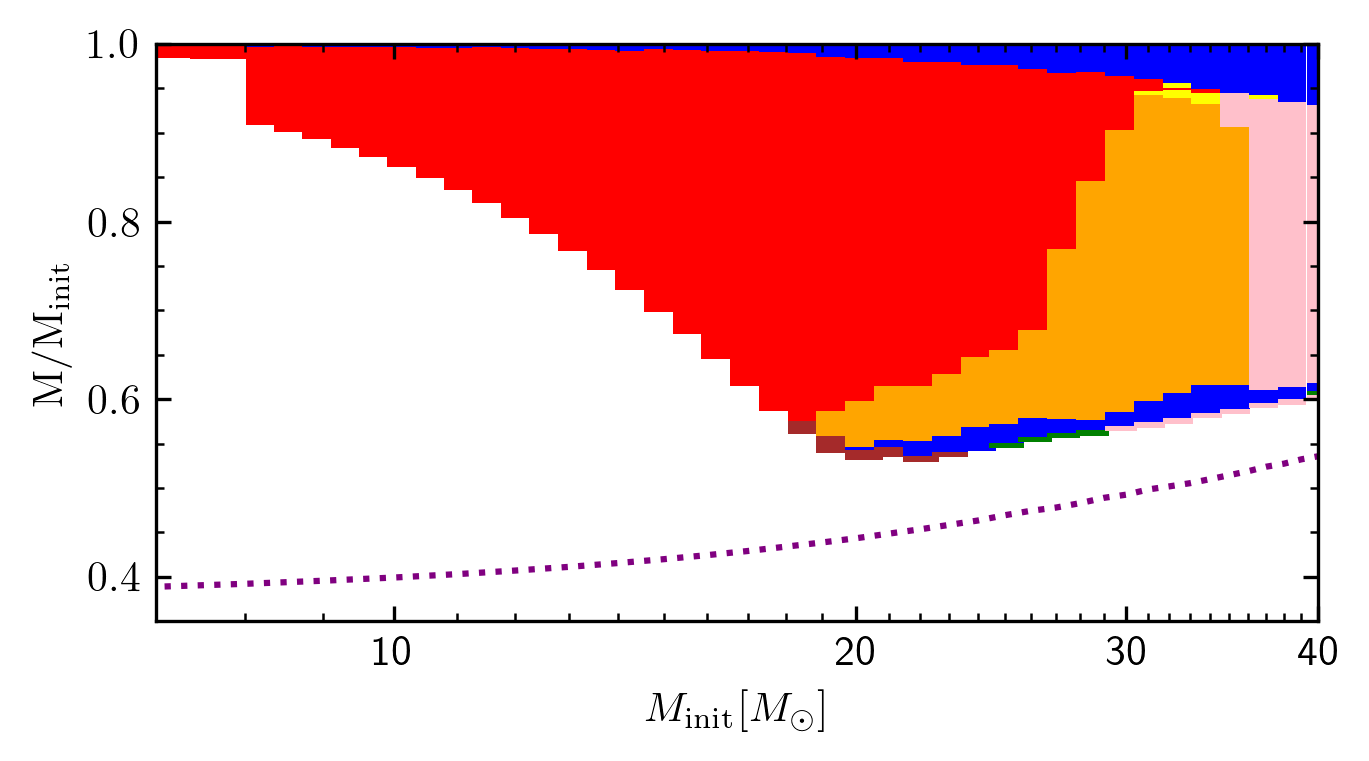}\\
\vspace*{-5.9pt}
\includegraphics[width=0.5\linewidth]{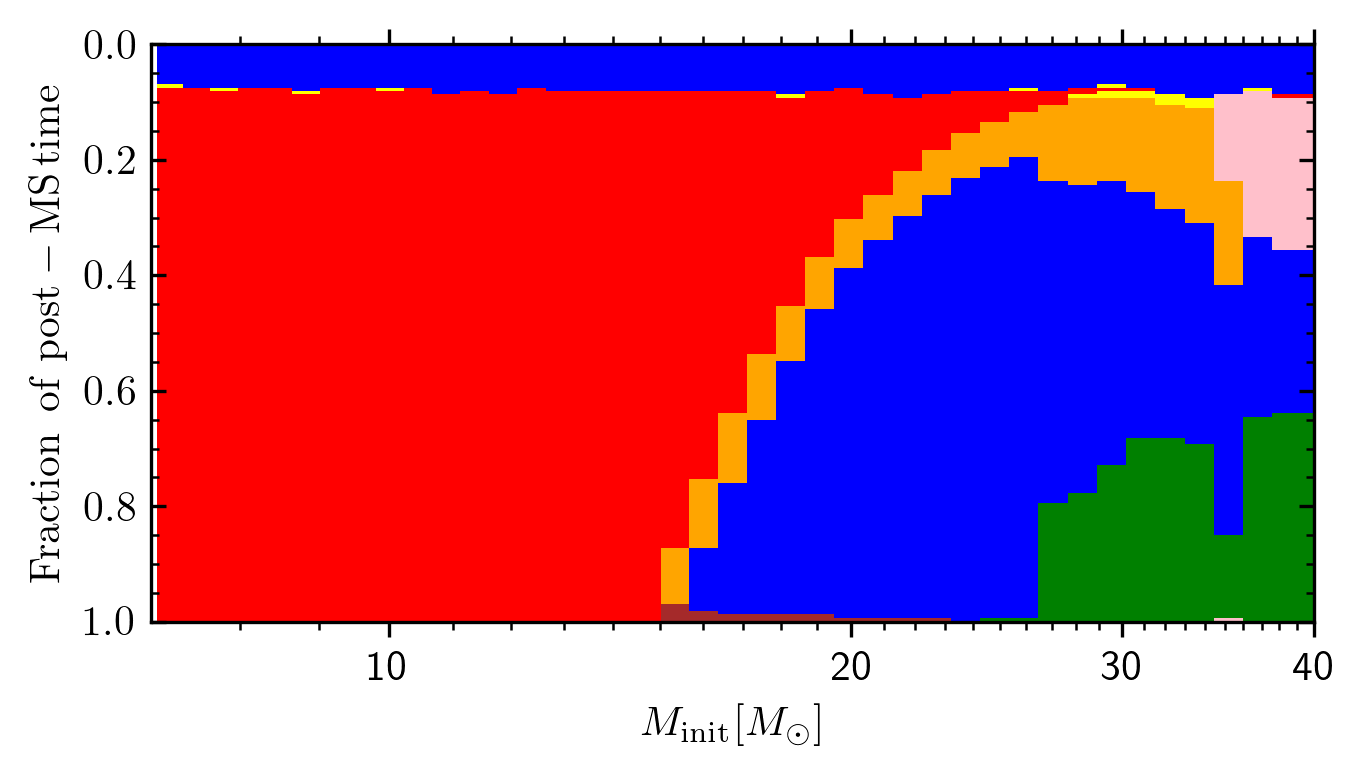}\includegraphics[width=0.5\linewidth]{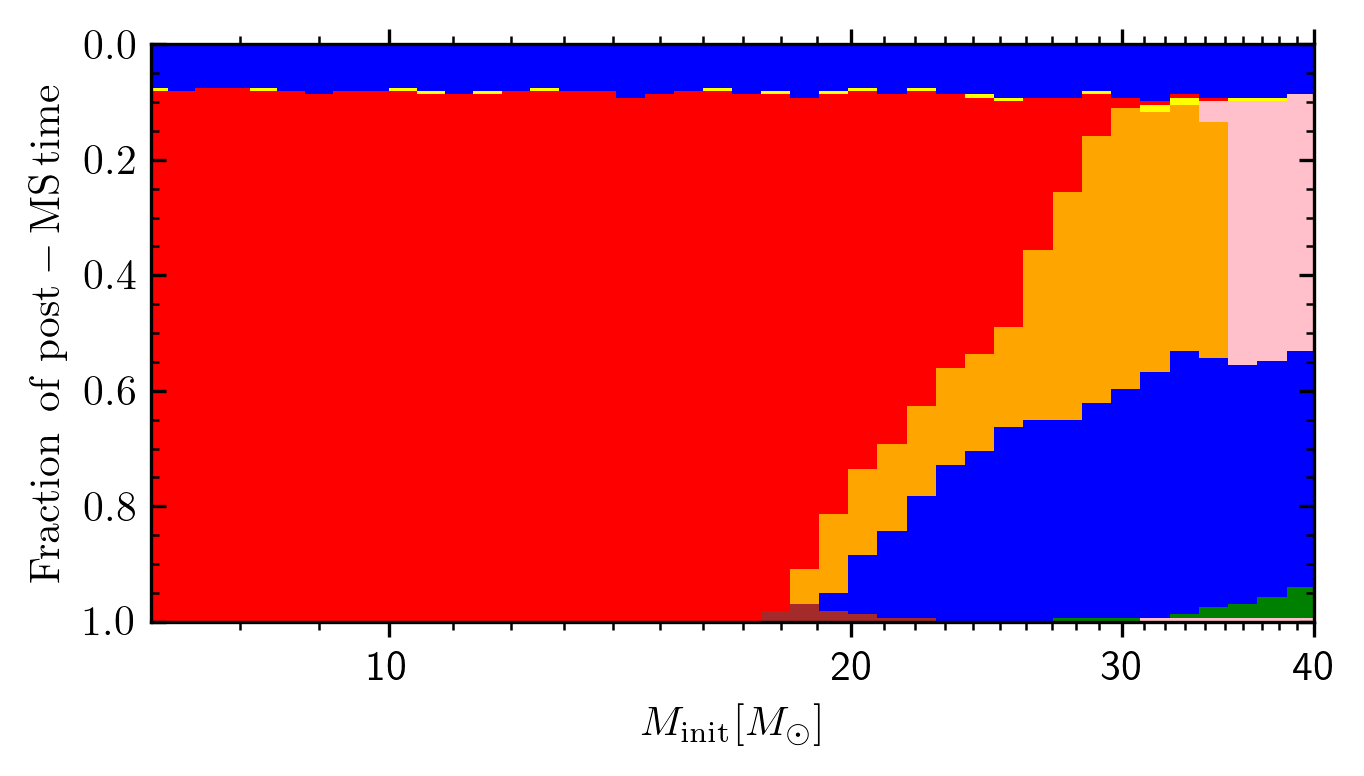}
\caption{Evolution during different phases. {\it Top}:  Indicative tracks of equal timesteps of 1kyr in the HRD with \yang (left) and \dejager (right), depicting the definition of each evolutionary phase. The initial mass is mentioned at the beginning of each track. The HD limit, where LBV-like winds kick in is also shown (grey dot-dashed line). {\it Middle}: Part of the stellar mass lost in each evolutionary phase, as a function of initial mass 
(evolution goes from top to bottom). We also show the value of the final helium core mass as a function of the initial (purple dotted line). {\it Bottom}: Fractional time spent in each evolutionary phase after the end of MS (TAMS), as a function of initial mass (evolution goes from top to bottom).
} 
\label{fig:Mlost_Time_Minit_andHRD}
\end{figure*}

\subsection{Observational sample of RSGs in the SMC}\label{sec:obs_sample}

  To test our population models, in Sect~\ref{sec:comparison_with_obs} we compare with the sample of RSGs in the SMC compiled by \yangt, which is based on the catalogs from \citet{Yang+2020} and \citet{Ren+2021}. 
  The collection, after excluding potential AGB contaminants, consists of 2,121 targets and comprises the largest, most complete 
  sample of RSGs in the SMC so far. The \yang prescription was derived based on this refined sample. 
  The effective temperature $T_{\rm eff}$ of the sources is inferred using the $J-K_{\rm S}$ relation of \citet[][Eq. 1]{Yang+2023}. For the dereddened colors, we assume a constant $A_{V, \rm CSM}=0.1$, which is the typical value found for most RSGs in that study. We also add a foreground Galactic and SMC internal extinction of $E(B-V) = 0.04$ and $0.05$ \citep{Massey+2007}, respectively, combined with a canonical $R_{\rm V}=2.74$ value for SMC \citep{Gordon+2003}, $(A_{J}/A_{V})_{\rm 2MASS} = 0.243$ and $(A_{K}/A_{V})_{\rm 2MASS} = 0.078$ \citep{Wang+2019}. We adopt a distance modulus of $18.95$ mag to the SMC 
\citep{Graczyk+2014, Scowcroft+2016}.  
  The luminosity of the sources has already been calculated in \citet{Yang+2023} by integrating the spectral energy distribution (SED) of each target, which is our default luminosity estimate. Alternatively, we also calculate a luminosity from the \citet{Neugent+2020} bolometric correction of the $K_{\rm S}$ magnitude of each source. Depending on these two luminosity calculation methods, we find 
  699 and 759 sources, respectively, 
  from the refined \yangt sample within our conservative limits for RSGs above $L_{\rm min}$ and below  $T_{\rm eff, max}$. As we demonstrate in Sect.~\ref{sec:comparison_with_obs}, although we include both luminosity calculations, their differences are not significant.


\begin{figure*}[t]
\centering
\includegraphics[width=0.5\linewidth]{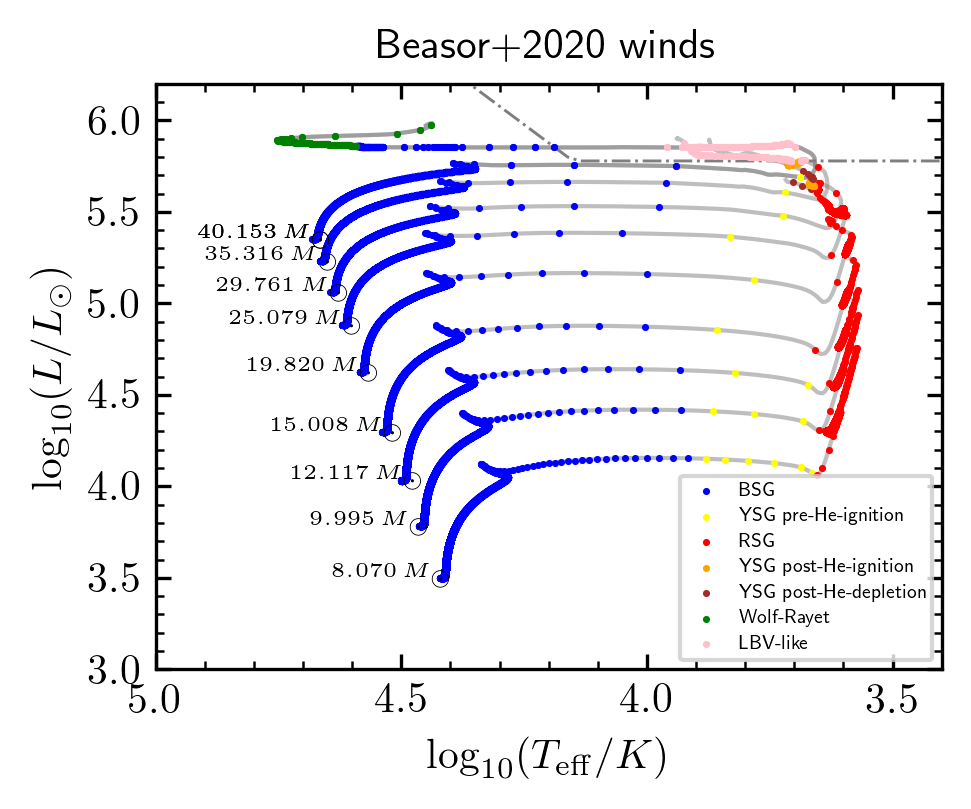}\includegraphics[width=0.5\linewidth]{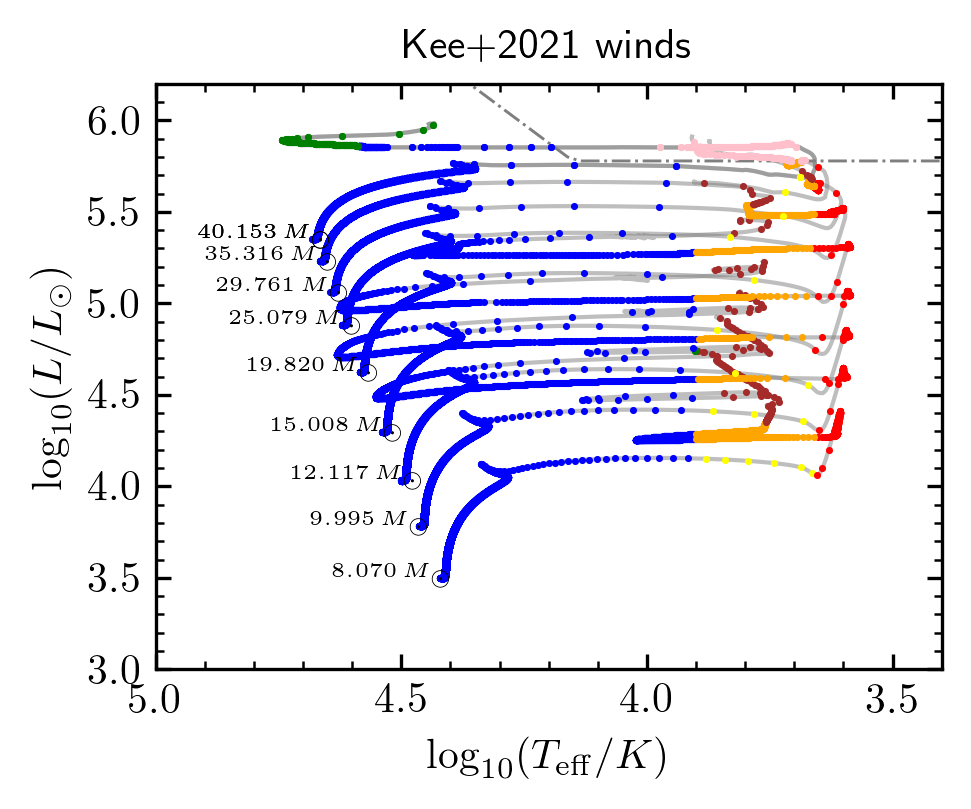}\\
\includegraphics[width=0.5\linewidth]{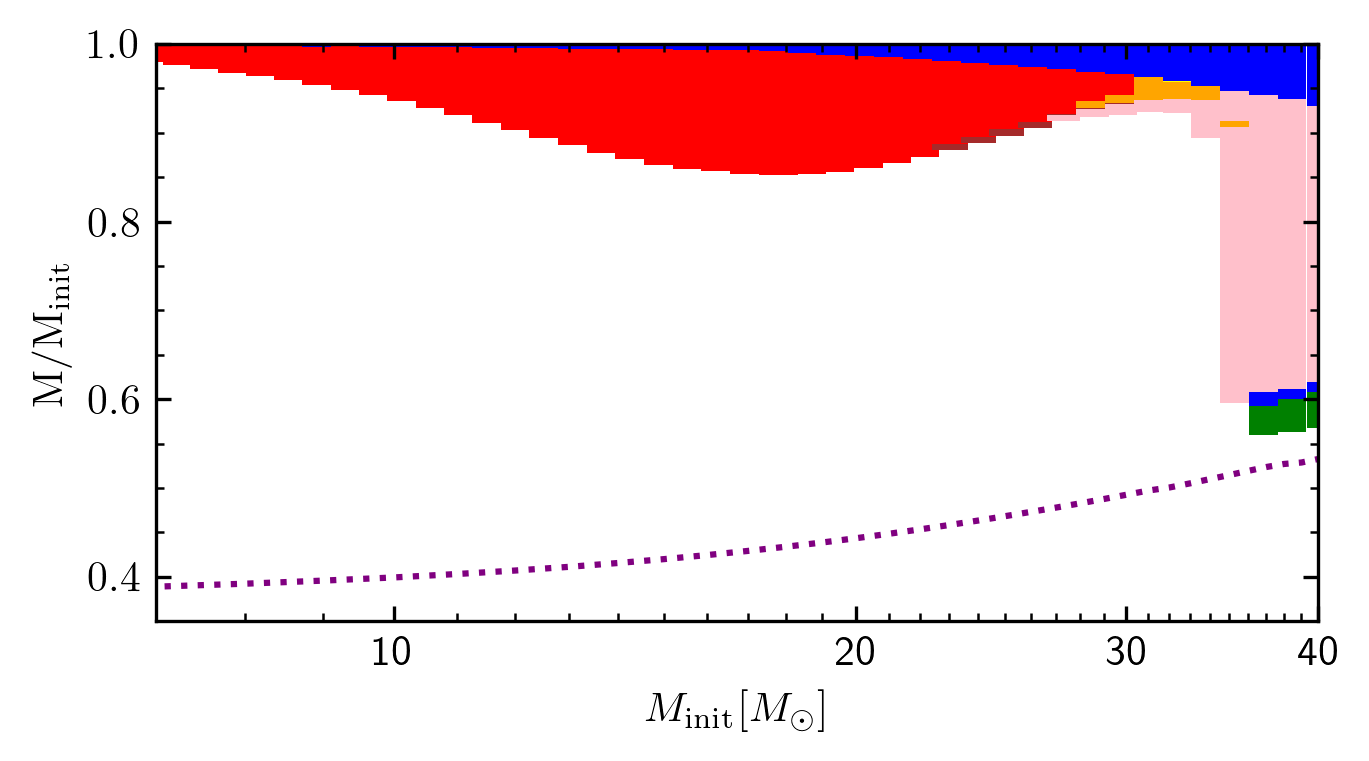}\includegraphics[width=0.5\linewidth]{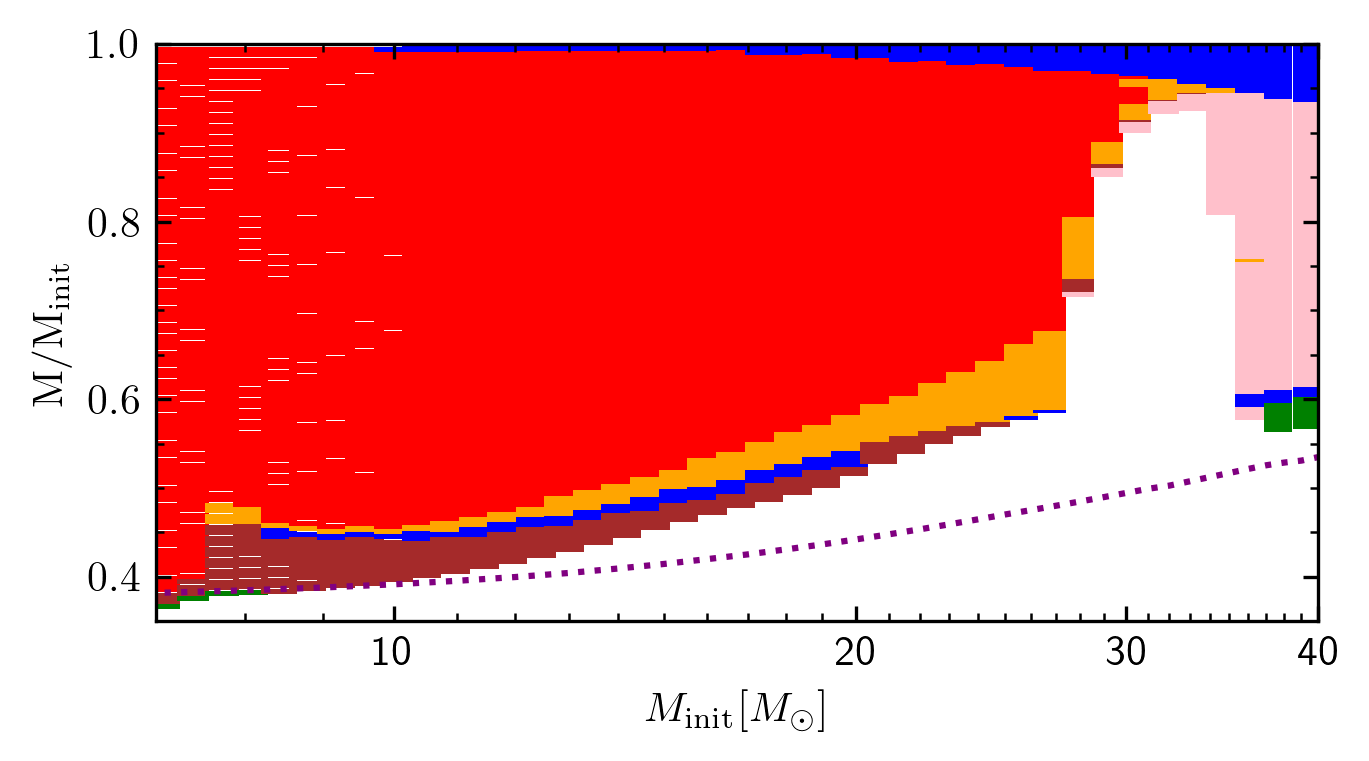}\\
\includegraphics[width=0.5\linewidth]{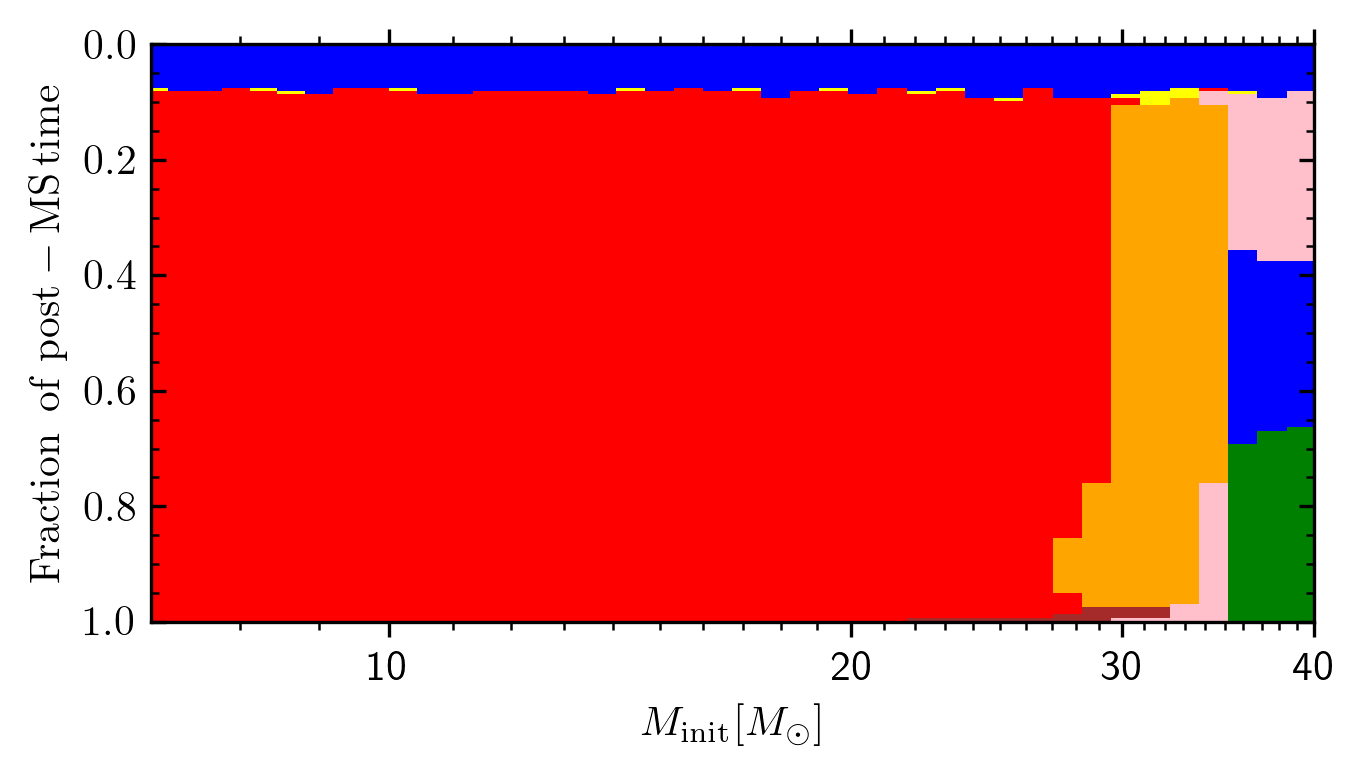}\includegraphics[width=0.5\linewidth]{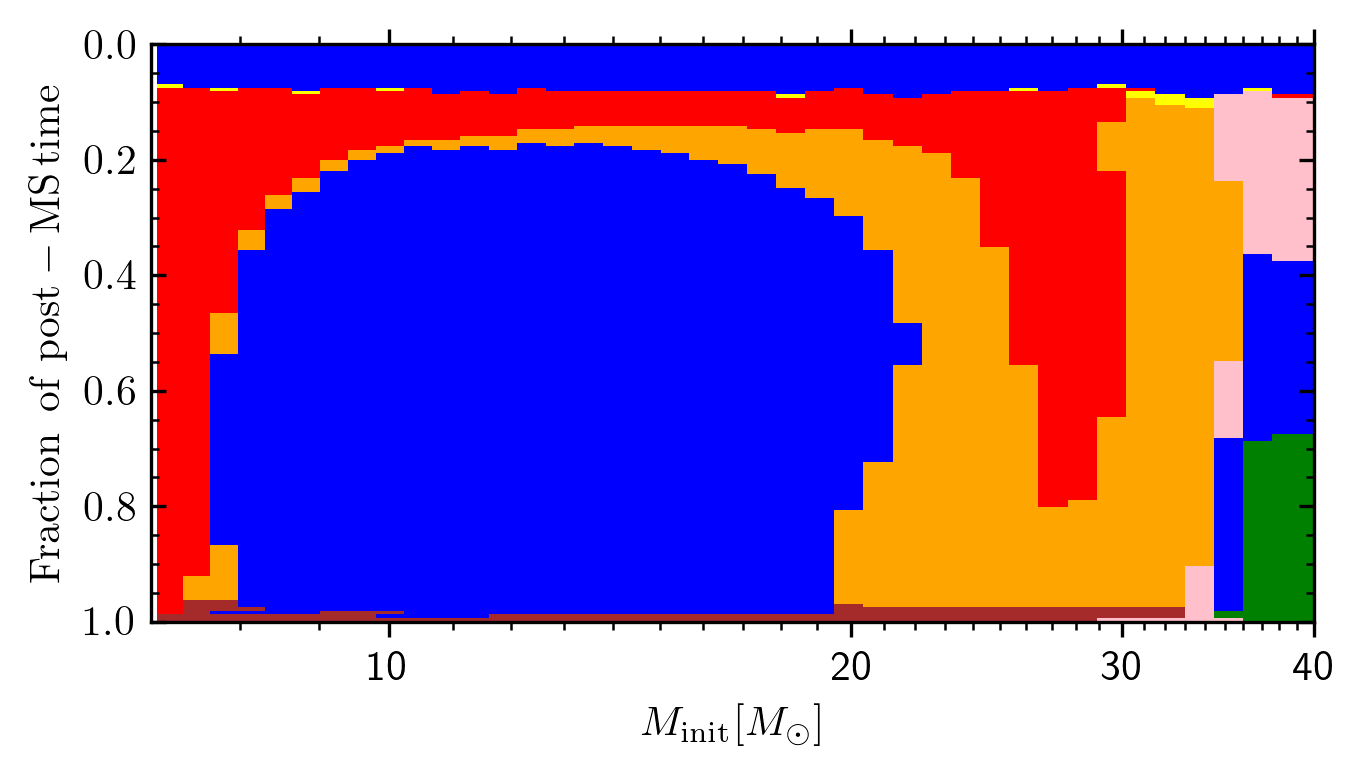}
\caption{Same as \fig{fig:Mlost_Time_Minit_andHRD} but for \beasor and \kee.
}
\label{fig:Mlost_Time_Minit_andHRD_appendix}
\end{figure*}

\section{Results}\label{sec:results}




\subsection{Mass lost and time spent during the RSG phase}\label{sec:mass_lost_time}

To understand the evolution of massive stars subject to the different RSG mass-loss prescriptions, we study the time they spend and the amount of mass they lose during their life, including their RSG phase. 
In the top panels of \fig{fig:Mlost_Time_Minit_andHRD} we see the evolution of a few indicative tracks on the HRD, colored by their evolutionary phase. Stars spend most of their time in the blue regime ($\log_{10}(T_{\rm eff}/K) > 3.9$) during their MS after which they expand quickly towards cooler temperatures. They pass  through a short-lived ``yellow'' phase \citep[$3.9 > \log_{10}(T_{\rm eff}/K) > 3.66$, following the definitions from][]{Georgy+2013, Meynet+2015} before initiating helium core burning (yellow), and promptly become RSGs ($\log_{10}(T_{\rm eff}/K) <3.66$).  

\begin{figure}
\centering
\includegraphics[width=1.\linewidth]
{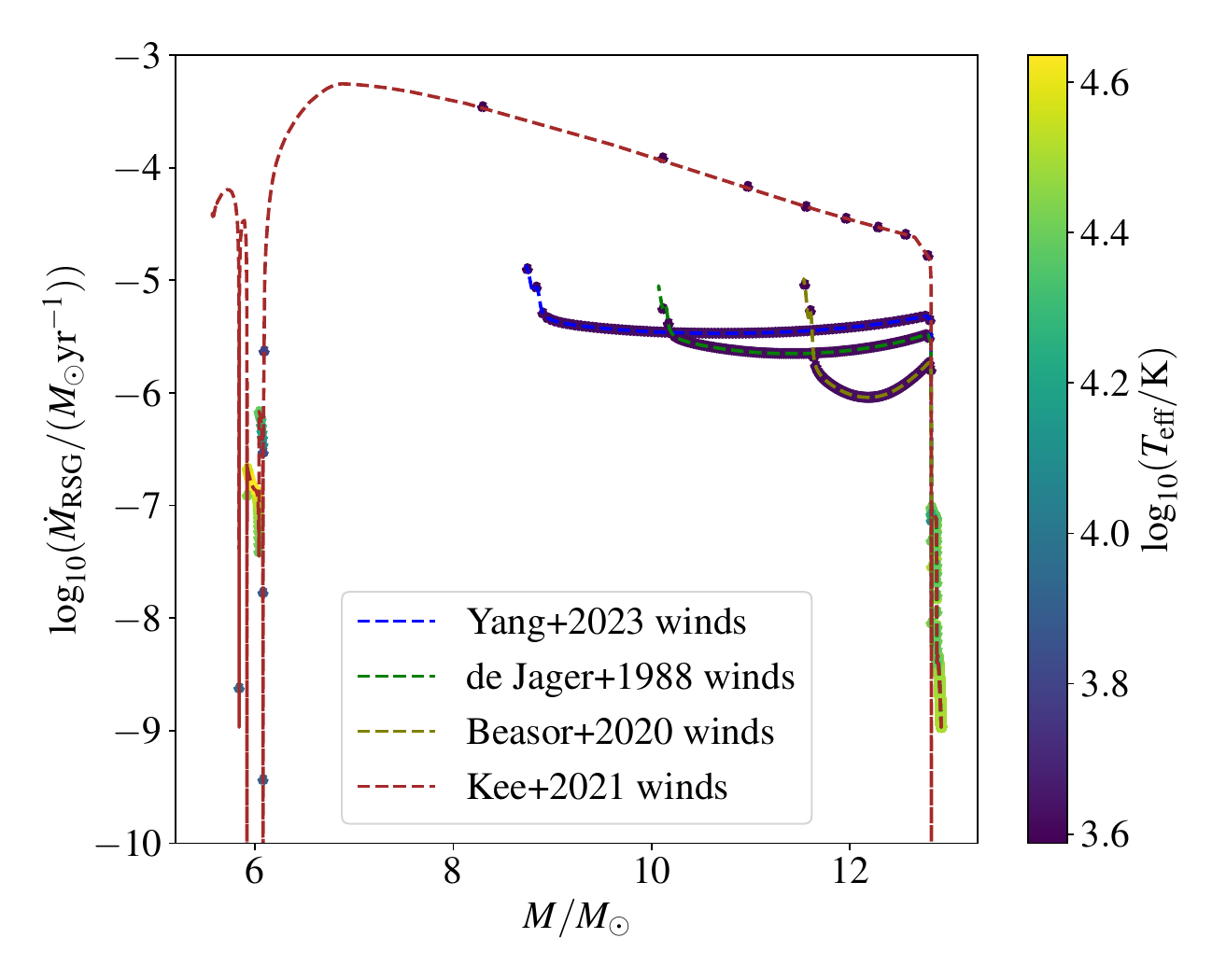}
\caption{The evolution of the mass-loss rate as a function of decreasing mass  (from right to left) is shown for an example track with an initial mass of $M_{\rm init} = 12.92 \Msun$, using the four different RSG mass loss prescriptions (see legend). Points are equally spaced timesteps of $10^4\,\ {\rm yr}$. The mass-loss rate increases when the $T_{\rm eff}$ (colorbar) drops after the MS, with \kee demonstrating the runaway effect of increasing mass-loss rate as the RSG decreases, becoming stripped and evolving bluewards in $\sim 7\times 10^4\,\ {\rm yr}$.}
\label{fig:M_Mdot}
\end{figure}

\begin{figure}[t] 
\centering
\includegraphics[width=0.99\linewidth]{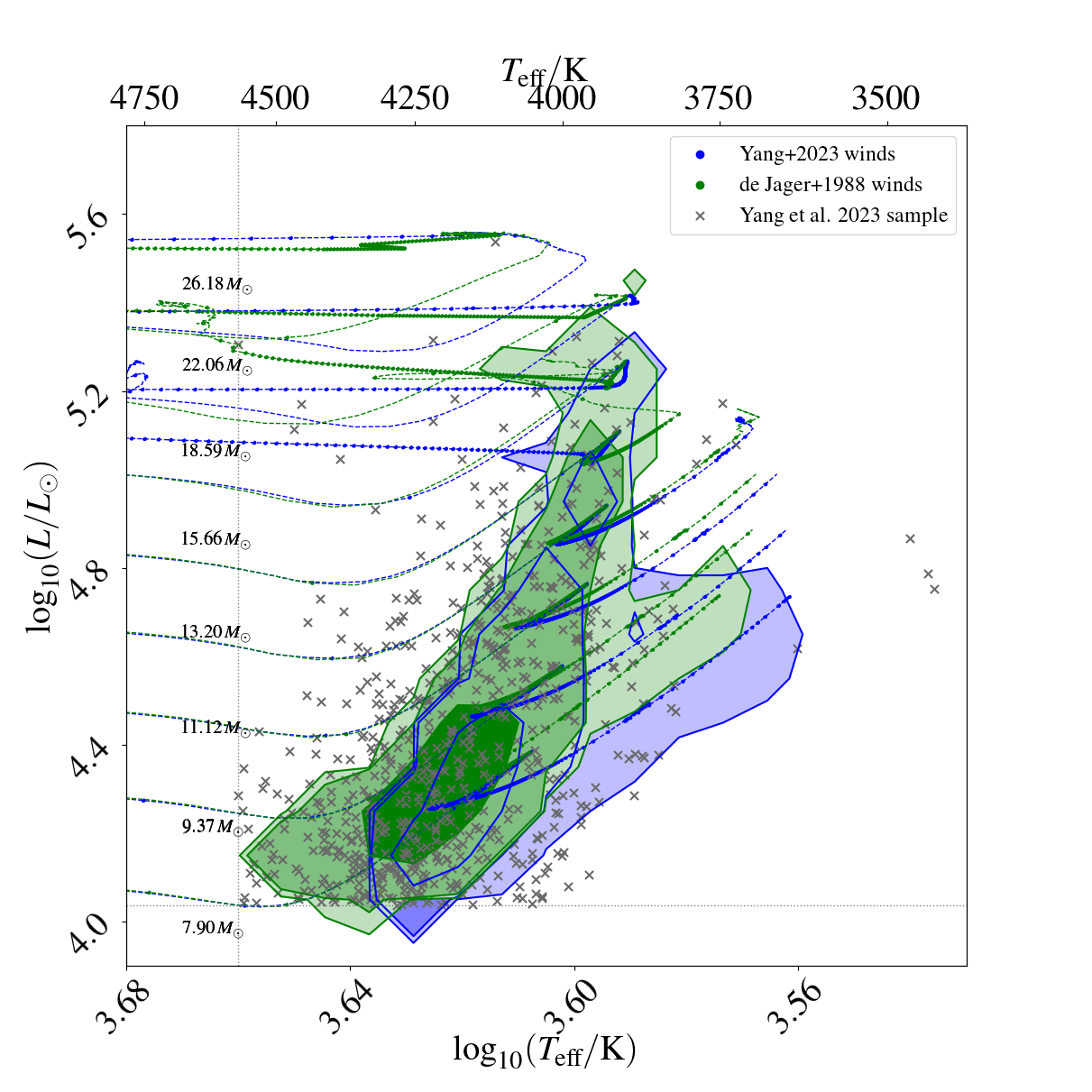}
\includegraphics[width=0.99\linewidth]{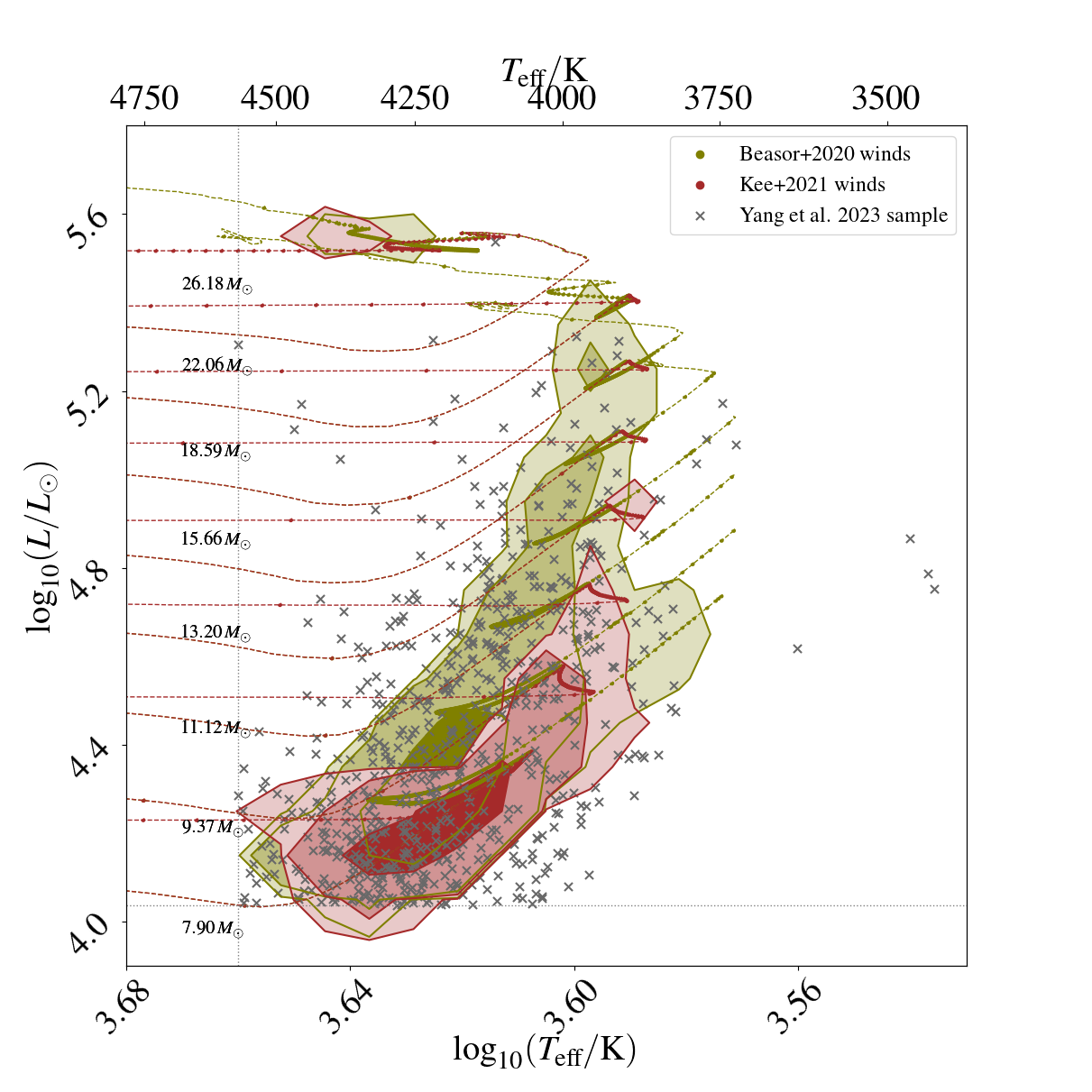}
\caption{
The Hertzsprung-Russell diagram of the RSGs in SMC. The contours show the probability density of the position of RSGs that lose mass according to \yang and \dejager (top panel) or \beasor and \kee (bottom), with contour levels denoting the 68\%, 95\%, and 99\% confidence regions. Representative stellar tracks for various initial masses are also displayed, with points corresponding to evenly-spaced timesteps of $2\times10^3$ years. 
Grey points depict the observed RSGs from \citet{Yang+2023} refined sample. We only show the results inside the conservative limits for a RSG of $L_{\rm min}$ and  $T_{\rm eff, max}$.
}
\label{fig:HRD_2D}
\end{figure}

Stars of $M_{\rm init} \lesssim 15$
following \yang (left column), spend almost all of their post-MS life in the RSG phase (bottom panels of \fig{fig:Mlost_Time_Minit_andHRD}), until collapse. The fractional mass lost during that period becomes a minimum for stars around $M_{\rm init} \sim  10 \Msun$, being around 30\% of the initial mass  (middle-left panel of \fig{fig:Mlost_Time_Minit_andHRD}). For our \posydon models and according to Eq.~\ref{eq:Minit_L}, this corresponds to $\log_{10}(L_{\rm RSG,avg}/\Lsun) \sim 4.6$, which is the location of the ``kink", i.e., the turning point of \yang towards even higher mass loss rates for increasing luminosity \citep[][also found in the RSG sample of the Large Magellanic Cloud; \citealt{{Antoniadis+2024}}]{Yang+2023}. 
Below this kink, \yang is almost constant with luminosity, and combined with the prolonged RSG lifetime for lower initial masses, leads to a higher fraction of their mass lost for them before \fecc. On the other hand, the increasing mass-loss rate above the kink is the reason for the curve-like relative mass loss, shown in the middle-left panel of \fig{fig:Mlost_Time_Minit_andHRD} for \yang. 
In contrast, for \dejager, the fraction of the initial mass lost to winds can be as low as $\sim 15\%$ for stars around $10\Msun$. This fraction becomes negligible for $M_{\rm init}<8\Msun$, where the \citet{Reimers1975} prescription kicks in. Conversely, this fraction increases to $\sim 50\%$ at higher initial masses.  
For all RSG mass-loss prescriptions, stars below $M_{\rm init} \sim 25 \Msun$, lose most of their mass during their RSG phase. This remains true even if they transition back to the yellow or blue regions of the HRD and spend a significant amount of their remaining lifetime there. 
This occurs either because the RSG mass-loss prescriptions result in a reduced mass loss at higher effective temperatures (e.g., \dejager, \kee) or because line-driven winds are weaker in the bluer parts of the HRD \citep{Vink+2000}.
The earlier departure from the RSG phase for stars with higher initial masses, along with the longer duration they spend in the yellow or blue regions during helium core burning, results in a slight reduction in the fractional mass loss.

RSGs originating from initial masses of approximately $15 \msun$  or greater, following \yang, tend to evolve blueward.
This happens because they lose a substantial portion of their envelope (as illustrated in the middle-left panel of \fig{fig:Mlost_Time_Minit_andHRD}), not because they undergo a "blue loop" which is a result of internal restructuring. 
They spend a considerable portion of their post-MS phase as helium-core burning yellow stars (orange points in \fig{fig:Mlost_Time_Minit_andHRD}) or once again in the blue region of the HRD, possibly becoming yellow again in their final stage after core helium depletion (brown  points in \fig{fig:Mlost_Time_Minit_andHRD}).
%
The transition where RSGs shift to the blue happens at initial masses as low as $15\Msun$ due to the higher mass loss rate of \yang above the turning point of the prescription, which increasingly diverges from \dejager by more than a factor of two at $\logL \sim 5.2$, and even more for more luminous RSGs. 
In contrast, for moderate mass-loss prescriptions, with a constant slope with luminosity, as in \dejager, the transition to blueward tracks occurs for stars with $M_{\rm init} \gtrsim 19-20\Msun$. 
Even then, massive stars spend a significant amount of their post-MS time in the yellow regime during helium core burning before going to the blue one (bottom-right panel). 

RSGs of $M_{\rm init} \gtrsim 27 \Msun$ get stripped and leave the RSG phase, even before helium core burning starts. They then go even more blueward, and when their surface hydrogen mass fraction drops below $0.4$, the Wolf-Rayet mass loss kicks in. Stars of $M_{\rm init}=27-40\Msun$ following \yang reach that stage earlier in their evolution, spending 20-35\% of their post-MS time as Wolf-Rayet stars.

In \fig{fig:Mlost_Time_Minit_andHRD_appendix} we plot the same information but for \beasor and \kee. Stars following the much weaker \beasor stay in the RSG phase (or just cross the $\log_{10}(T_{\rm eff}/{\rm K})=3.66$ limit and become yellow for stars of $M_{\rm init} \sim28-34 \Msun$) until the end of their evolution. They lose $\lesssim 15\%$ of their total mass during their whole evolution, except for stars with initial masses $\gtrsim 34\Msun$ that enter the HD limit  where enhanced LBV-like winds begin to strip the star down to its helium core, pushing it towards the blue region. 
In contrast, the RSG mass-loss rate of models following \kee is even stronger than in \yang 
for the moderate luminosities typical of most RSGs.
This is because \kee induces a runaway effect: as a star enters the RSG phase, it experiences significant mass loss, which reduces its surface gravity and mass. Since \kee is highly sensitive to these parameters, this leads to an even more intense wind, increasing by orders of magnitude, until the star is stripped enough to shift to the blue region, even for stars with an initial mass lower than $10\Msun$. 
In \fig{fig:M_Mdot}, we illustrate this runaway effect with an example system, showing that for a star of approximately $\sim 12.9 \msun$, its mass-loss rate increases from $\log_{10}(\dot{M}_{\rm RSG}/{\msun {\rm yr}^{-1}}) \approx -5$  upon entering the RSG phase to about $\sim -3.5$ as it becomes progressively stripped after just  $\sim 7\times 10^4\,\ {\rm yr}$, with only $6\msun$ remaining when it leaves the RSG phase moving bluewards. In comparison, the mass-loss rate following the other prescriptions remains around $\log_{10}(\dot{M}_{\rm RSG}/{\msun {\rm yr}^{-1}}) \approx -6$, $-5$, with the star remaining in the RSG phase until the end of its evolution.
This runaway behavior of \kee has been found in other model tests \citep[][Kee \& de Koter, priv. comm.]{Kee2024}.  
On the other hand, for initial masses between  $27-34\Msun$, the mass loss of \kee when it enters the RSG regime is even lower than \dejager (see \fig{fig:Mdot_luminosity} for $\logL \sim 5.6$), preventing this runaway effect from occurring. 
In this part of the parameter space, the total mass lost with \kee is relatively low, around 10-20\% of the initial mass. 
However, for stars with even higher initial masses, they become sufficiently luminous to reach the HD limit, where their total mass lost increases again due to LBV-like winds.

%

%






\subsection{Comparison with observations of RSGs}\label{sec:comparison_with_obs}

To compare with observations, we simulate populations of massive single stars with \posydon, using four grids of stellar tracks each implementing a different RSG wind. In \fig{fig:HRD_2D} we show the expected position in the HRD of the population of RSGs. As discussed in Sect.~\ref{sec:simulation_SMC}, the contours are weighted by the IMF, the star formation history of SMC, as well as the time a RSG spends in each position in the HRD. 
Some indicative evolutionary tracks are also shown (one every eighth track on the grid). 
We also show the 
674 RSG from the refined observational sample of \yangt (grey points) that are inside the limits of  $\log_{10}(L_{\rm min}/\Lsun) \sim 4.03$ and below $\log_{10}(T_{\rm eff, max}/K) = 3.66$. 
%


\begin{figure}
\centering
\includegraphics[width=\linewidth]%
{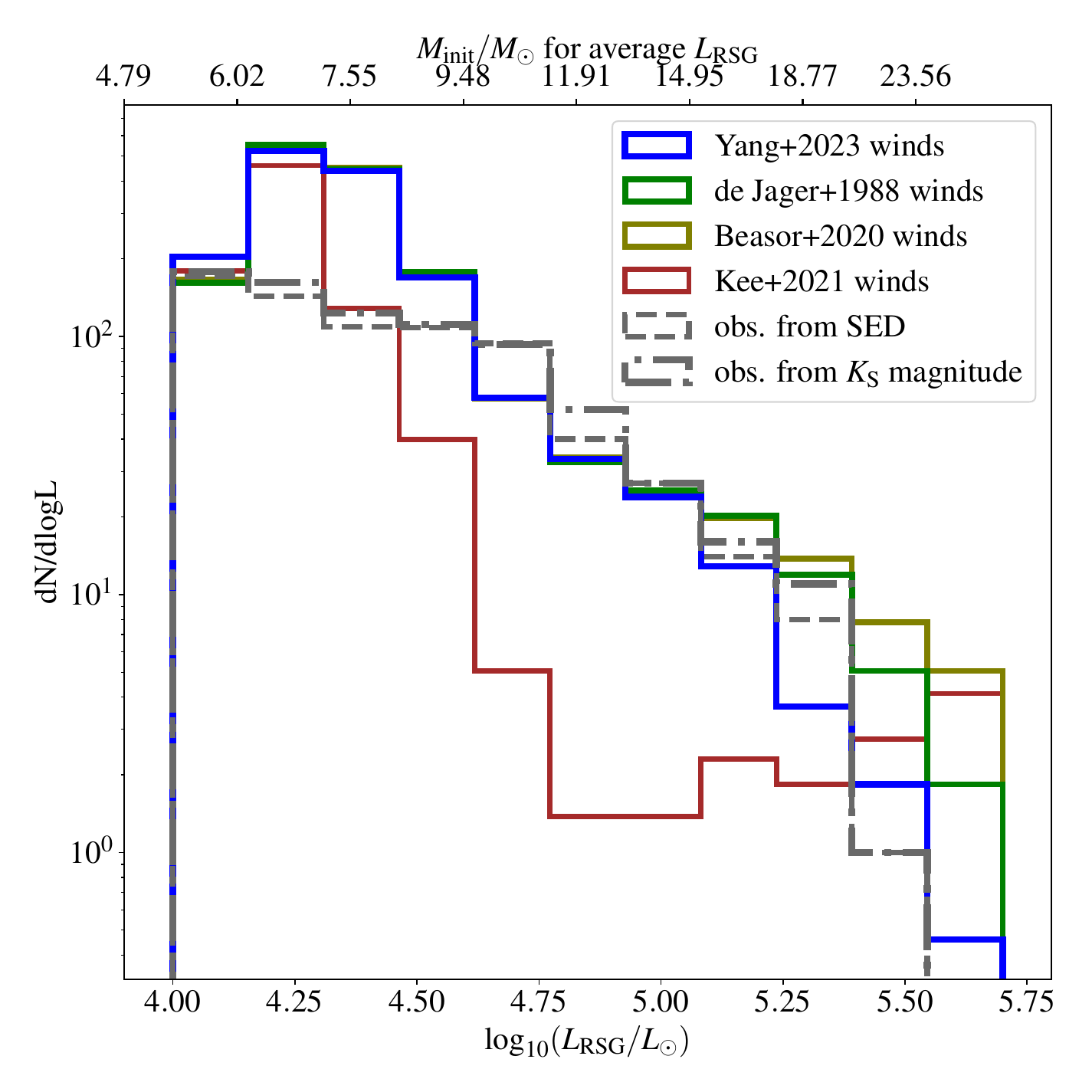}
\caption{Theoretical and observed luminosity functions of RSGs from \yangt. 
We limit our comparison to systems that eventually result in a \fecc, and for which $\log_{10}(L_{\rm min}/\Lsun) > L_{\rm min}$, and $T_{\rm eff} < T_{\rm eff,max}$. 
The top $x$-axis shows the corresponding initial mass of models with an average RSG luminosity of the bottom $x$-axis, according to Eq.~\ref{eq:Minit_L} following the best-fit parameters for \yang. 
}
\label{fig:Luminosity_function}
\end{figure}

Although the mass lost from a star during its RSG phase varies significantly depending on the assumed prescriptions, RSGs 
will occupy a similar position in the HRD for a vast range 
 of envelope masses \citep[e.g.,][]{Justham+2014,Farrell+2020b}. 
The key difference from the various RSG mass loss prescriptions pertains to how long they remain as RSGs, as they become partially stripped of their H-rich envelope and evolve bluewards. 
The higher the mass-loss rate, the earlier in their life a RSG of the same initial mass will become stripped (see bottom row of \fig{fig:Mlost_Time_Minit_andHRD} and \ref{fig:Mlost_Time_Minit_andHRD_appendix}). Furthermore, as the mass-loss rate is highly sensitive to the luminosity, we expect luminous RSGs (originating on average from initially more massive stars) to be more affected by the assumed wind prescription \citep[][]{Neugent+2020}. 
To depict the outcome of this effect on the population, and to further quantify the comparison with observations, we compute the luminosity function of RSGs in \fig{fig:Luminosity_function}. The modelled and observed distributions of the effective temperatures and the mass-loss rates can be found in \fig{fig:Teff_Mdot_distr}. 

The luminosity function indirectly probes the (average) mass-loss rate of the RSGs during their lifetimes, not the instantaneous one \citep{Massey+2023}, as it is affected by the total mass lost from a RSG and whether it is sufficient to go bluewards. This tool remains agnostic to the mechanisms of mass loss and thus can indirectly probe possible episodic mass loss episodes throughout the RSG evolution.
Thus, a stronger mass-loss rate is expected to lead to steeper luminosity functions of RSGs, because the luminous ones would leave the RSG phase even earlier.  
Indeed, we find that models with high RSG mass-loss rates, such as \kee and even \yang have a steeper decline towards high luminosities compared to \dejager or \beasor.

The observed number of the RSGs and the shape of the luminosity function are consistent with the predicted ones from \dejager, \yang and \beasor, for $4.6 \lesssim \log_{10}L \lesssim 5.0 $. 
On the high luminosity side, \yang causes the predicted luminosity function to drop slightly earlier than the observed one, in contrast to the \dejager and \kee that result in a shallower decline, although we are in the regime of very low statistics. This is due to the ``kink'' in \yang prescription with even higher mass-loss rates for the luminous RSGs, opposite to \beasor.  
The luminosity function is depleted of luminous RSGs when we assume \kee and vastly deviates from the others and the empirical one. Still, a few very luminous RSG around $\logL \sim 5.6$ persist, originating from stars of $M_{\rm init} \sim 27\Msun$ where a decrease in total mass lost is predicted, as discussed in Sect.~\ref{sec:mass_lost_time}.

We note that the luminosity functions in \fig{fig:Luminosity_function} are not normalized, instead they depict the expected (and observed) number of RSGs per luminosity bin. 
For all the models, apart from \kee, we find $\sim 1500$ expected RSGs above $L_{\rm min}$. The predicted numbers are in tension with the 699 RSGs in the observed refined sample of \citet{Yang+2023}. This number can increase to $759$, if we use the bolometric correction of the $K_{\rm S}$-magnitude for the luminosity estimate. 
In both cases, we assume a constant $A_{V \mathrm{,CSM}}=0.1$ (an average of \citealt{Yang+2023} for all targets). Of course, our theoretical expectations of the total number of RSGs are sensitive to the exact shape of the IMF and the star formation history (SFH) of the SMC. The formation and interaction of binary systems would also play a role, and we further discuss this in Sec~\ref{sec:binarity}. 
Although the population following \kee predicts $\sim 850$ RSGs, which seems more consistent with the observed number, as we see below this prescription cannot match the distribution of luminosities.

The observed RSGs are much fewer than predicted at low luminosities with $\logL \lesssim 4.6 $ (\fig{fig:Luminosity_function}). The latter may be due to the decreasing completeness of the sample for lower RSG luminosities, even for $\logL>L_{\rm min}$. 
\citet{Yang+2020,Yang+2023} point out that fainter RSGs are often missed by photometric surveys. Even when observed, they may have less reliable photometry and can be rejected as having poor quality.
Furthermore, blended or unresolved sources may exist, especially in crowded regions. \yangt applied photometric quality criteria and cuts to avoid AGB contamination removing around 211 sources from their initial sample, which may include some rare dust-enshrouded RSGs \citep{Beasor+2022}, reddening their colors to resemble photometric AGBs. 

\begin{figure}
\centering
\includegraphics[width=0.9\linewidth]{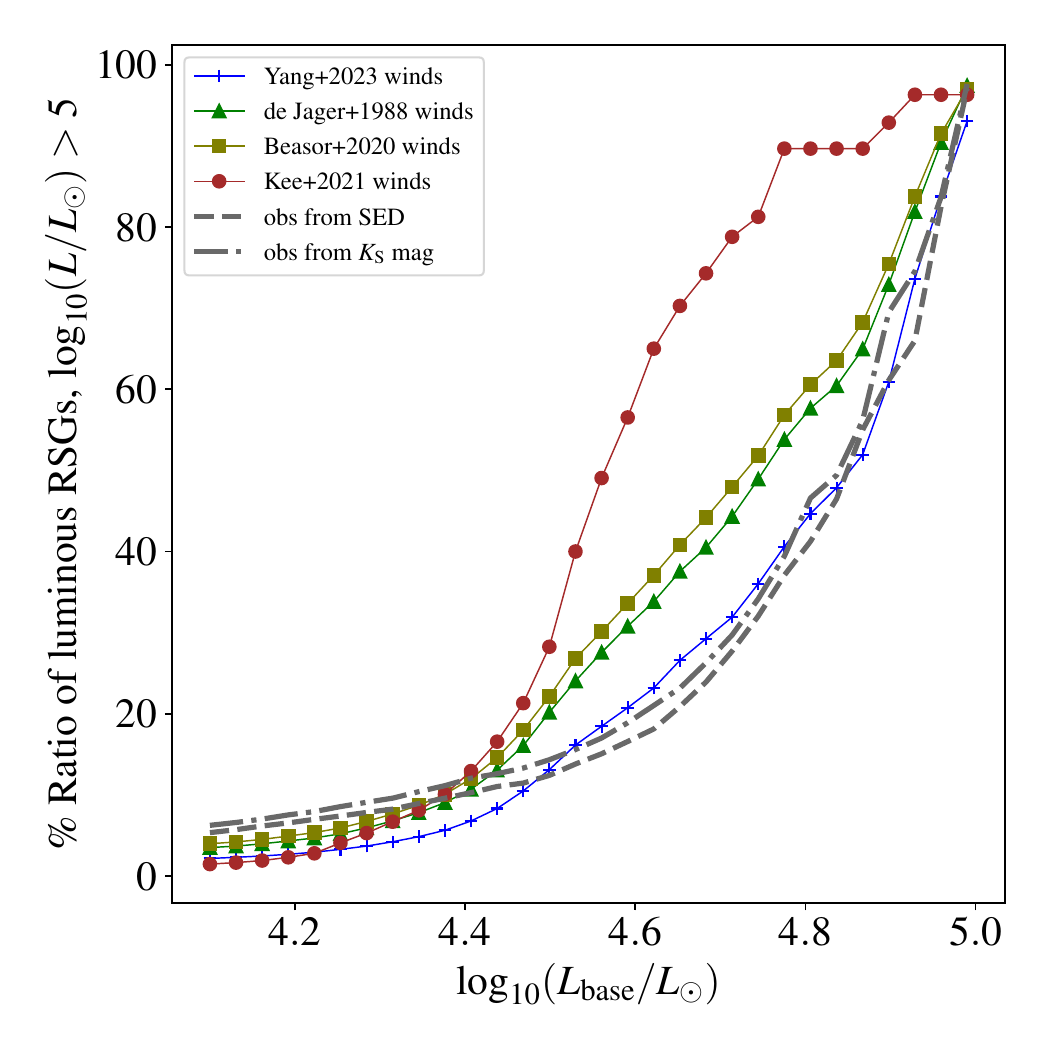}
    \caption{Ratio of luminous RSGs ($\log_{10}(L/ \rm L_{\odot}) > 5.0$) to all RSGs above a variable base luminosity, $L_{\rm base}$, shown in the x-axis.}
\label{fig:Ratio_luminous}
\end{figure}

We try to take into account possible bias against less luminous RSGs and any potential issues with the normalization of the luminosity function, by showing in \fig{fig:Ratio_luminous}  the ratio of luminous RSGs with $\log_{10}(L/ \rm L_{\odot}) > 5.0$ divided by all RSGs above a base luminosity, $L_{\rm base}$, that we vary, as in \citet{Massey+2023}. For higher base luminosities,  the ratio by definition increases and we expect the completeness of the observed sample to also increase. The observed ratio goes from a few percent to $\sim 15\%$ for $\log_{10}(L_{\rm base}/ \rm L_{\odot}) = 4.5$. \kee has a sharp increase of the predicted ratio as we go to higher base luminosities, dominated by the few very luminous predicted sources in combination with the depletion of the moderate luminous ones, which are quickly stripped as they reach the RSG phase. Both \beasor and \dejager seem to overestimate the number of luminous RSGs due to their weak mass loss, which retains them in this evolutionary phase without going bluewards. 
In contrast, \yang appears to be consistent with this empirical ratio, especially for $\log_{10}(L_{\rm base}/ \rm L_{\odot}) \gtrsim 4.5$, where the predicted excess of low-luminosity RSGs is not taken into account.

\subsection{RSGs as Type II supernova  progenitors}\label{sec:SNprogenitors}

\begin{figure}
\centering
\includegraphics[width=1.05\linewidth]{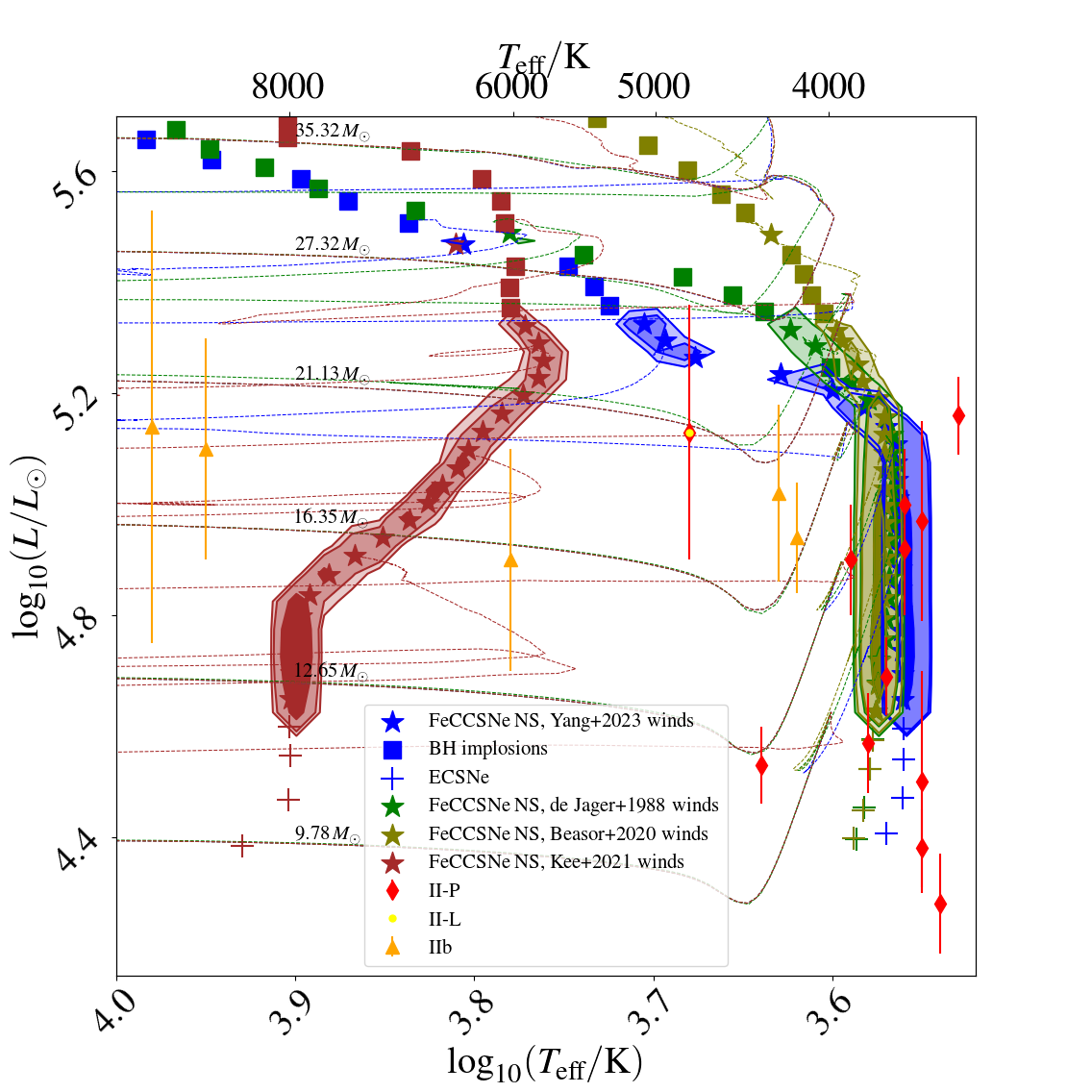} 
\caption{HRD with the expected positions of the final state of the stars, before core-collapse. Successful \fecc explosions for each track \citep[according to the SN prescription by][]{Patton+2020} are depicted with a star, implosions into a BH with probably no transient with a square, and \ecsne with plus signs. Contours represent the positions of the \fecc events only, 
with different colors representing the different RSG mass loss prescriptions. We also show the position of the detected type II SN progenitors from the compilation of \citet[][]{Farrell+2020a}. 
}
\label{fig:RSG_SNprogenitors}
\end{figure}

The different RSG mass loss mechanisms will also affect the outcome of massive stars. 
%
%
In \fig{fig:RSG_SNprogenitors} we depict the region in the HRD that we expect the SN progenitors with $T_{\rm eff}< 10^4~{\rm K}$  to lie for the different RSG mass loss assumptions. Our models reach technically up to carbon core depletion but we expect no significant change in the stellar global properties and thus in their HRD position during the remaining years to decades until the actual collapse, assuming no obscuration from extra pre-SN mass loss \citep[e.g.,][]{Davies+2022}. The 2D histogram takes into account the weighted-with-SFH$_{\rm SMC}$ IMF of the stellar progenitors, but in contrast to \fig{fig:HRD_2D}, it is not influenced by the time spent by them in their RSG phase. It only takes into account the final  position of massive stars that are expected to lead to successful \fecc explosions according to the \citet{Patton+2020} prescription (based on the progenitor's carbon-oxygen core mass and central carbon $^{12}\mathrm{C}$ at the end of helium core depletion), which predicts some ``windows" in the initial mass parameter space in which progenitors directly implode into a black hole. Thus, it excludes core-collapse events that do not result in observable transients due to black-hole implosion (shown with squares). In addition, as discussed in Sect.~\ref{sec:simulation_SMC} we exclude from the analysis progenitors that are expected to produce \ecsne, although they are still shown in  \fig{fig:RSG_SNprogenitors} with plus signs. 
%
The observed HRD positions of nearby Type II-P, II-L, and IIb SN progenitors are overplotted for comparison \citep{Smartt+2015}. We also include a few indicative low-luminosity SN progenitors \citep[from the sample shown in][]{Farrell+2020a}, which are thought to arise from low-mass RSG progenitors \citep[e.g.,][]{O'Neill+2021,Valerin+2022} and may be the progenitors of low-energetic \ecsne events \citep[e.g.,][]{Lisakov+2018}.

\beasor and \dejager both form RSG progenitors that would be consistent with type II SN observations. 
For even higher luminosities,  they predict hotter progenitors, becoming even YSGs for \dejager. 
However, most of them are presumed to not explode due to their high core masses, according to \citet{Patton+2020}. Thus,  both prescriptions reproduce the lack of observed, luminous, high-mass RSG SN progenitors, introduced as the so-called ``Red-supergiant problem''.

\fecc progenitors following \yang are also consistent with the positions of observed ones up to  $\log_{10}(L/\Lsun) \sim 5.2$. The models diverge towards bluer final HRD positions compared to \dejager and \beasor, due to more severe stripping as we go to higher final luminosities. Still, most of them are expected to not explode  according to \citet{Patton+2020}, predicting hotter candidates for failed \fecc compared to \dejager and \beasor. Thus, \yang models are also consistent with the RSG problem. 
Few of these YSGs final models predicted by \yang around that luminosity turn, are expected to explode, but have not been observed yet. Although the positions in the HRD of the exploding YSGs according to \yang are close to the detected Type II-L SN2009kr progenitor, it is debatable whether this source is the core-collapse progenitor \citep{Maund+2015}.  The predicted  temperature range of the YSG progenitors ($\sim$4000--5200~K) may be consistent with the observed Type IIb progenitors, but their predicted luminosity of $ \log_{10}(L/\Lsun) \gtrsim  5.2$ is $0.2-0.3$ dex higher.  
Note also that predicted RSG progenitor positions, below  $\log_{10}(L/\Lsun) \sim 5.1$, are found slightly hotter by $\lesssim 200$~K following \beasor compared to \dejager and especially \yang. This is due to the higher envelope mass at the moment of collapse \citep[see bottom figure 2 of][]{Farrell+2020a}.

The high mass loss stripping found in our models following \kee  is inconsistent with the formation of Type II-P and II-L SN progenitors, although it may lead to Type IIb SNe, due to the thin hydrogen-rich envelope these stars still have at the end of our simulations.  However, the majority of Type IIb  SNe may originate from alternative binary stripping scenarios \citep[e.g.,][]{Nomoto+1995, Claeys+2011, Yoon+2017, Sravan+2020}, as in the case of SN 2008ax \citep{Folatelli+2015} and SN 1993J \citep{Podsiadlowski+1993,Nomoto+1993}, for which candidate binary companions have been suggested \citep{Maund+2004,Fox+2014}. 
Possible single-star progenitors are expected to originate from a very small range of masses, as fine-tuning is needed for a thin hydrogen-rich envelope to be left, and even then it is not clear which is the fraction that undergoes direct collapse onto a black hole \citep{Zapartas+2021b}.


\section{Discussion}\label{sec:discussion}

\subsection{Possible binary history of red supergiants}\label{sec:binarity}

\begin{figure}
\centering
\includegraphics[width=1.\linewidth]{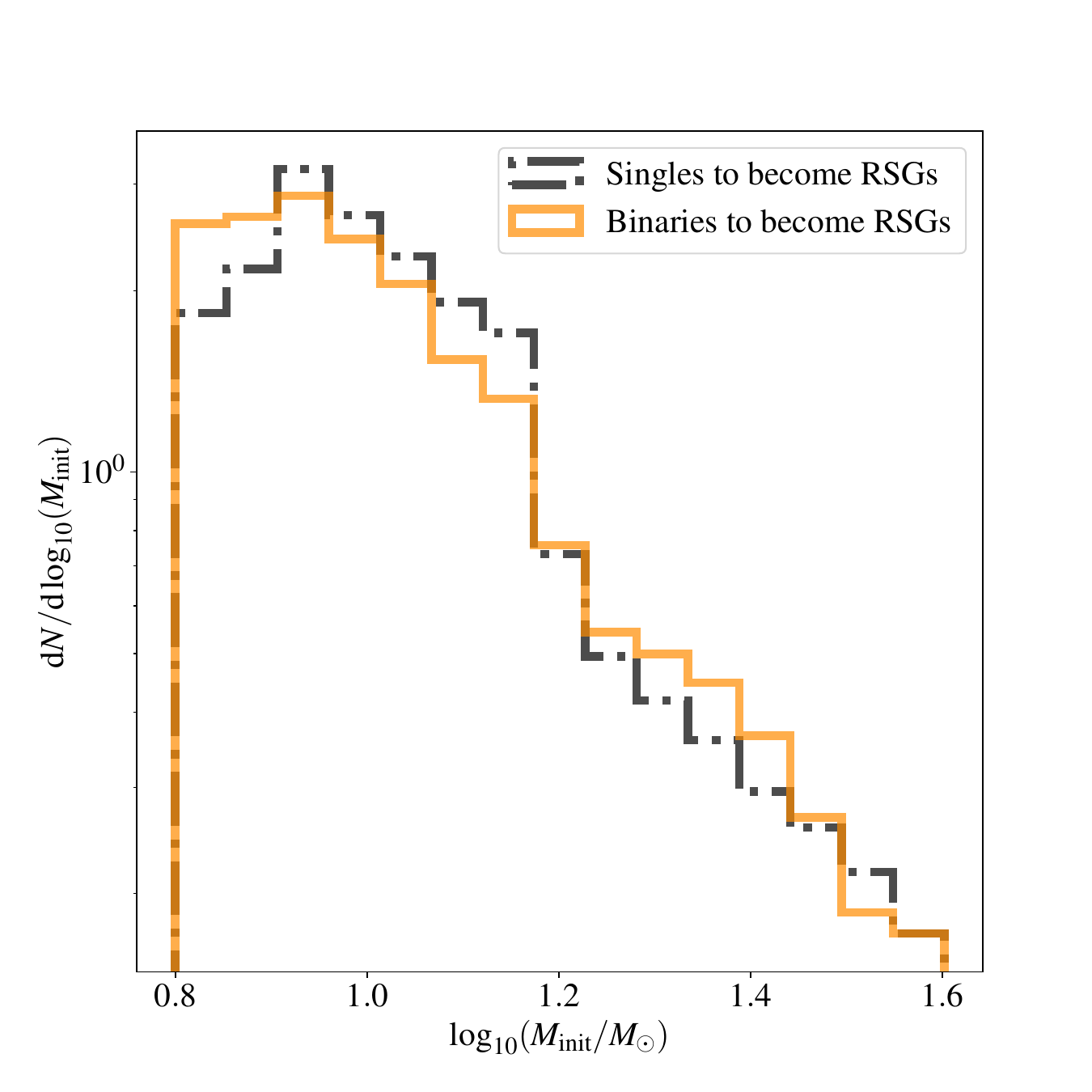} 
\caption{Normalized, SFH-weighted IMF for massive stars that reach TAMS and eventually will become RSGs, for a single star population (grey) and from one where binarity is taken into account (orange). For binaries that at some point become RSGs, $M_{\rm init}$ is calculated based on their $M_{\rm TAMS}$ (Eq.~\ref{eq:TAMS_ZAMS_mass}), taking into account possible accretion (or merging) during their MS.
}
\label{fig:bin_IMF}
\end{figure}

\begin{figure}
\centering
\includegraphics[width=1.\linewidth]{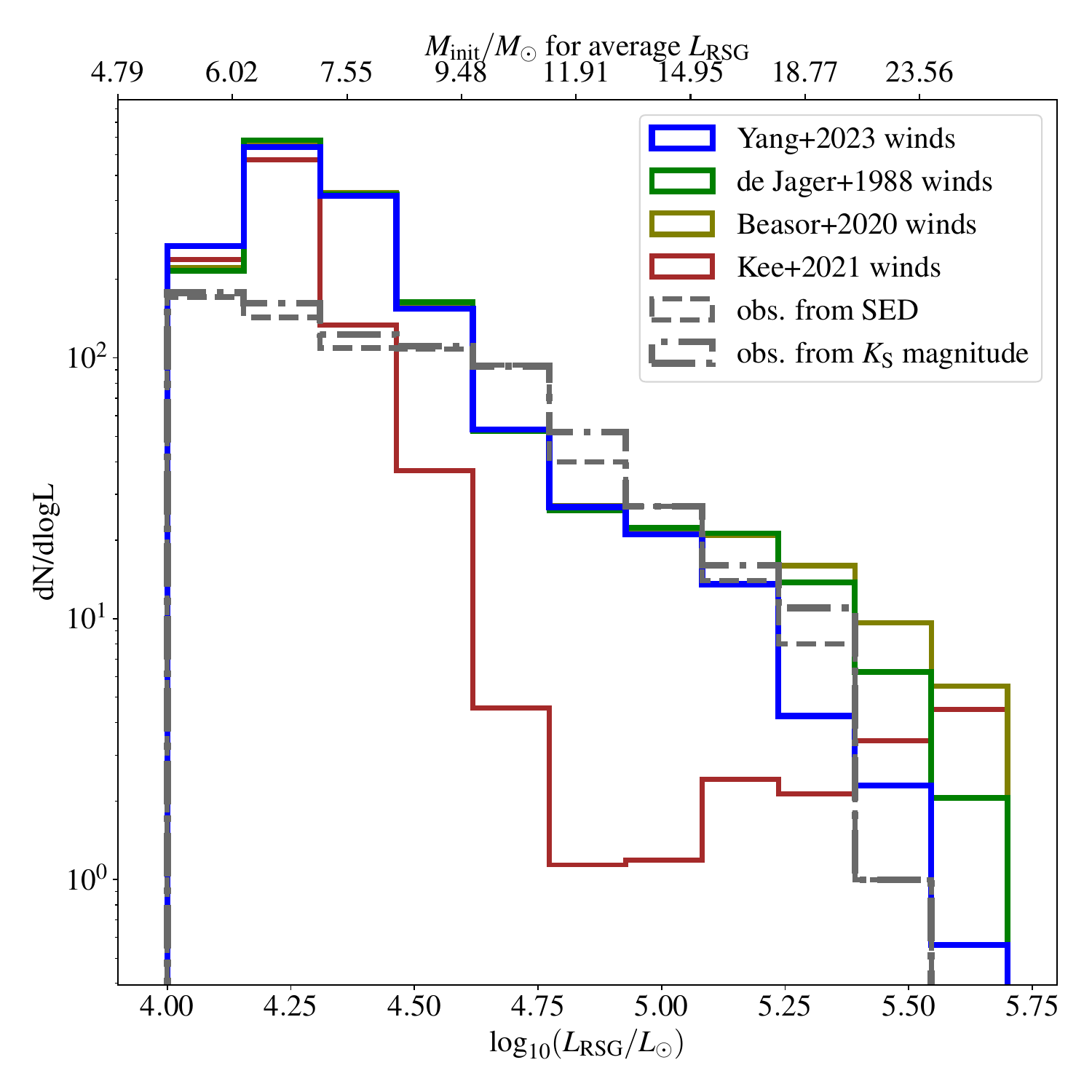} 
\caption{Same as \fig{fig:Luminosity_function}, but with a binary-corrected relative IMF for RSG progenitors, as shown in \fig{fig:bin_IMF}.
}
\label{fig:bin_Lum_function}
\end{figure}

\begin{figure}
\centering
\includegraphics[width=1.\linewidth]{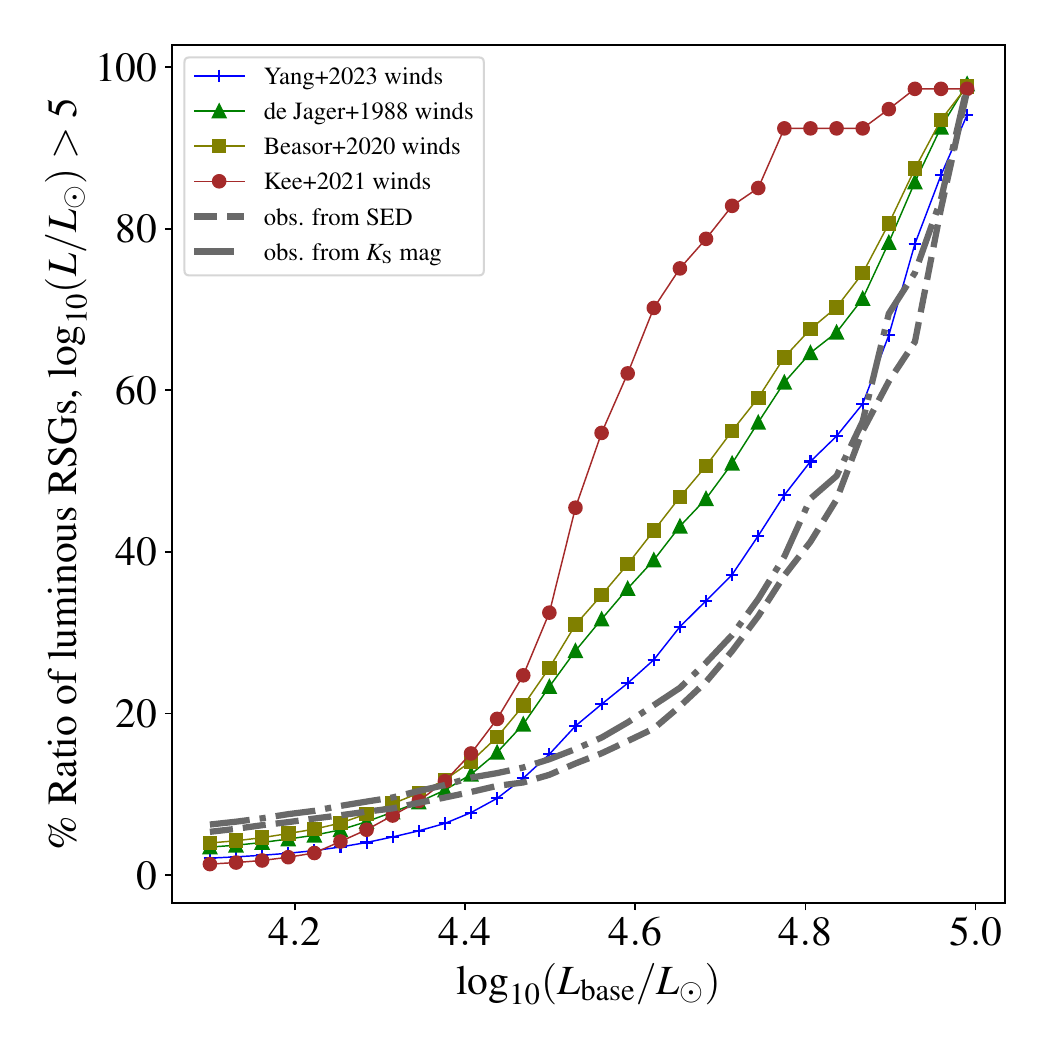} 
\caption{Same as \fig{fig:Ratio_luminous}, but with a binary-corrected relative IMF for RSG progenitors, as shown in \fig{fig:bin_IMF}.}
\label{fig:bin_Ratio_luminous}
\end{figure}

In our analysis so far, we have overlooked the impact of binary interactions, even though most massive stars are born in binary or multiple systems, with a significant fraction of these binaries interacting during their lifetimes and initiating mass transfer \citep[e.g.,][]{Sana+2012,Almeida+2017}. As a result of mass transfer, stars can be stripped of their hydrogen envelopes \citep[e.g.,][]{Paczyski1971}, which are then accreted by the companion stars, or the binary system may merge. 
In the former case, donor stars stripped through mass transfer might not experience a RSG phase \citep[e.g.,][]{Gotberg+2017}, impacting the overall shape and distribution of the RSG luminosity function. At the same time, the mass-gaining stars will form a more massive core and thus experience stronger winds due to their increased luminosity, affecting their subsequent evolution and stellar properties during the RSG phase. In the latter case, stellar mergers can also influence the RSG luminosity function as they can alter both the number of stars that undergo a RSG phase,  and the evolution of the merger product \citep[e.g.,][]{Chatzopoulos+2020, Menon+2024}.

Indeed, it is now established that binary history can explain many observed features of the red supergiant population, including runaway red supergiants  \citep[mass gainers ejected from a prior SN;][]{Comeron+2020} and the presence of red supergiant stragglers \citep{Britavskiy+2019a}. In addition, according to \citet{de-Mink+2014}, about 15\% of the observed massive blue stars are predicted to be merger products, with many of them expected to become RSG later in their evolution. 
This is further supported by studies that predict a significant fraction of binary products in the population of RSG that are Type II SN progenitors \citep{Podsiadlowski+1992,Zapartas+2019}, including mass-gaining stars, mergers of two main-sequence stars (MS+MS), or mergers involving at least one post-main-sequence star \citep[postMS merger, e.g.][]{Blagorodnova+2021}.

Here, we aim to provide a rough estimate of how these different binary evolution scenarios influence the luminosity function of RSGs, by evolving a population of massive binary systems according to \posydon v2 \citep{Andrews+inprep}. 
Knowing that RSG luminosity primarily depends on the helium core mass formed at the end of the MS phase \citep[TAMS, defined in our models when hydrogen central mass fraction drops below $1\%$][]{Farrell+2020b, Schneider+2024}, we use the mass of each binary product at that point, $M_{\rm TAMS}$, as an indicator of its eventual post-MS helium core mass formed and thus its $L_{\rm RSG}$.  This approach takes into account the increase in mass during the MS of the star (through MS+MS mergers or accretion onto MS stars scenarios),  
because a high level of mixing is expected for them due to the absence of a significant chemical boundary between the core and the formed envelope, and thus a higher helium core mass during its eventual RSG phase \citep[e.g.,][]{Braun+Langer1995,  Glebbeek+2013, Schneider+2024}. On the other hand, we ignore further mass accretion after TAMS, which is not expected to significantly increase the core mass any further. So, for RSGs resulting from postMS merging, we treat the merger products as single stars with the same core mass as the giant component. We also exclude donor stars that become stripped of their envelopes after expanding beyond the TAMS and thus never become RSGs. 
Specifically, we exclude those that do not reach a minimum effective temperature of  $\log_{10}T_{\rm eff, min} = 3.66$ at any point  during their life. In this calculation, we neglect any potential binary stripping during the RSG  phase. 
%
Our exploratory estimate eliminates the need for new stellar models that simulate the complex stellar structures expected in postMS mergers.  However, accurately incorporating merger models for the other two scenarios, which involve one or two evolved stars, requires a detailed simulation of merger products \citep[e.g.,][]{Menon+2024,Schneider+2024}. This is beyond the scope of this study and will be addressed in future work.

%

We assign to each remaining binary product a corresponding single-star initial mass using a fitting relation between  $M_{\rm TAMS}$ and $M_{\rm init}$ of \posydon single-star models (Eq.~\ref{eq:TAMS_ZAMS_mass}). In practice, we rearrange the IMF to approximate the effect of binarity on the luminosity distribution of its RSGs. 
To further weight each star with the SFH of the SMC, we assume that the time needed for the binary product to reach TAMS is also the time of its RSG phase (as the evolution from TAMS to the RSG phase is negligible and the duration of the RSG phase is short relative to the total stellar lifetime). The latter assumption is an acceptable approximation as the RSG timescale is much smaller than the bins in the SFH distribution we used as described in Sect.~\ref{sec:method}. Thus, a star that increased in mass during MS due to binary interaction  (and still became RSG) would correspond to a higher initial mass IMF, but will be assigned to the weight of the SFH according to its TAMS time.

 With the above calculation, we can effectively estimate the SFH-weighted IMF for future RSGs from binary products, relative to a population of RSGs from only single stars (\fig{fig:bin_IMF}). 
%
The peak of both distributions for $0.95\lesssim\log_{10}(M_{\rm init}/\Msun)\lesssim 1.2$ is due to the weighting of SMC SFH which boosts stellar sources with ages of 14-40\Myrs (Sect.~\ref{sec:method}).  Binary stars that reach TAMS and are expected to produce a RSG broaden the relative SFH-weighted IMF. This happens due to a boost at low initial masses of $\log_{10}(M_{\rm init}/\Msun)\sim 0.9$ from actually intermediate-mass stars that, by accreting or merging, they become massive enough to explode. Most of these intermediate-mass progenitors are binary accretors that become massive enough at TAMS to become RSGs eventually. At the same time, mass accretion or MS+MS merging to higher masses before TAMS leads to a rearrangement of the relative IMF, slightly decreasing the contribution around the peak towards higher initial masses  $\log_{10}(M_{\rm init}/\Msun) \gtrsim 1.2$. This is consistent with the slight shift of the default distribution of final core masses towards more massive ones compared to a canonical IMF, found in \citet[][Fig. 5]{Zapartas+2021a}.  

The impact of this relative change of the binary-corrected IMF of stars that eventually become RSGs, leads to a slightly different RSG luminosity function, seen in \fig{fig:bin_Lum_function}. The general shapes stay similar although the boost above $\log_{10}(M_{\rm initial}/\Msun) \gtrsim 1.2$ leads to a slight increase in the fraction of luminous RSGs for all assumed RSG mass loss prescriptions (\fig{fig:bin_Ratio_luminous}).  
%
This does not help in reconciling the discrepancy of the ratio according to \kee, \beasor and \dejager. \yang still stays closer to the observed ratio for different $L_{\rm base}$ values, especially at higher base luminosities (going to the right of this plot) where the ratio should be less  biased to the completeness of the sample. At the same time, binarity cannot reconcile the discrepancy with observations at the low-luminosity regime. 


\subsection{Predictions for post-RSG stars}
Large spectroscopic samples of
RSGs and YSGs are scarce. For empirical estimates of the YSG-to-RSG ratio (YSG:RSG), we typically rely on color-magnitude diagrams \citep[e.g.][]{Yang+2021} and machine learning techniques applied on the larger, photometric samples \citep[e.g.][]{Maravelias+2022, Dorn-Wallenstein+2023}.
In a sample of 4000 evolved massive stars, \citet{Yang+2021} estimated their YSG:RSG $\sim 17 \%$. The machine-learning approach of \citet{Maravelias+2022}, trained on the colors of the stars, predicts a YSG:RSG of $2-8\%$. In our models, we find a YSG:RSG of $\sim$0.8$\%$, $\sim$0.8$\%$, $\sim$0.7$\%$ and $\sim$14.6$\%$ using \yang, \dejager, \beasor and \kee respectively. The ratio due to the  high level of stripping from \kee is closer to the empirical values, although including the effects of stripping in binaries \citep[an efficient scenario to form YSG Type IIb SN progenitors][]{Yoon+2017,Sravan+2018} or of episodic mass loss when reaching the ionization bump below $\log_{10}(T_{\rm eff}/{\rm K}) \sim 4$ \citep{Cheng+2024} may drastically increase the theoretical ratio.

The stellar properties of YSGs that are formed after their RSG phase significantly differ from their pre-RSG counterparts \citep{DornWallenstein+2022, Humphreys+2023}, as a result of changes in the stellar core and subsequent mixing of material to the surface (for instance dredge ups), and mass loss through stellar winds affecting the mass and surface gravity. Assuming post-RSG YSGs are those showing short-period pulsations, \citet{DornWallenstein+2022} suggest that 33$\%$ of the YSGs above $\log_{10}(L/\Lsun) \sim 5.0$ are post-RSGs. Similarly, based on their expected infrared excess \citet{Gordon+2016} found 30-40$\%$ of the YSG population to be post-RSG. From the evolutionary tracks, we find the ratio of post-RSG yellow supergiants above $\log_{10}(L/\Lsun) \sim 5.0$ compared to all YSG to be  $\sim$30$\%$, $\sim$14$\%$, $\sim$4$\%$, and $\sim$23$\%$ following \yang, \dejager, \beasor and \kee respectively. Thus again, the degree of stripping predicted by the \yang best explains the observed pre-to-post YSG ratio, neglecting again other mass-loss mechanisms such as binary stripping or episodic mass loss.

\subsection{Tension with observations and caveats}

The implemented RSG mass loss prescriptions can reproduce various features of the tail of the luminosity function and the ratio of luminous RSGs, but are inconsistent with other aspects (Sect.~\ref{sec:comparison_with_obs}). In general, the high mass-loss rate of \kee appears inconsistent with the observed luminosity function, 
at least given the fixed set of the other stellar physical assumptions of \posydon (discussed below). 
Note that severe stripping following \kee would occur only for higher initial masses, for a different model-dependent $M_{\rm init}-L_{\rm RSG}$ relation (see the difference in the luminosities in our models compared to the ones used by \citet{Kee+2021} in Fig. A1).  Equivalently, a $v_{\rm turb}$ value lower than $18.2 \,{\rm km/s}$ would have resulted in weaker stripping, as the prescription is highly sensitive to this parameter.
%

The predicted mass-loss rates especially at high luminosities significantly affect the timespan of the RSG phase, thus the occurrence rate of luminous RSGs. The curved relation for RSG lifetime with initial mass for \yang (bottom left panel of \fig{fig:Mlost_Time_Minit_andHRD}) originates due to the change of the slope in \yang prescription with luminosity, with even higher mass-loss rates for the luminous RSGs (\fig{fig:Mdot_luminosity}). This ``kink" has been described in \yangt, and is also found in the more recent \citet{Antoniadis+2024} for the Large Magellanic Cloud. An effectively opposite change in slope is found in \beasor, due to the dependence on initial mass, thus indirectly on RSG luminosity. The effect of this is that 
\dejager and especially \beasor slightly overpredict the relative ratio of luminous RSGs (\fig{fig:Ratio_luminous}). These discrepancies seem to worsen when considering the effects of possible binary interactions of stars that eventually become RSGs (Sect.~\ref{sec:binarity}). A possible shallower IMF for SMC \citep[e.g.,][]{Schneider+2018} that increases the formation rate of massive stars would increase the ratio too, making the discrepancy even worse. On the other hand, \yang, which is based on the \yangt observed RSG sample that we compare with, seems to be the most consistent with the ratio of luminous RSGs. %
Interestingly, the lack of very luminous RSG due to \yang stripping seems to also naturally reproduce the HD limit \citep{Humphreys+Davidson1979} that has been updated around $\logL\sim 5.5$ for the Magellanic Clouds \citep{Davies+2018b} and M31 \citep{McDonald+2022}, without invoking an extra mass loss mechanism (which in our models is triggered only at $\logL\sim 5.78$, see Sec.~\ref{sec:Mdot_prescriptions}). 
In addition, its high mass-loss rates would produce more Wolf-Rayet stars stripped by winds, consistent with recent observations at the SMC \citep{Schootemeijer+2018,Schootemeijer+2024}. %

In contrast, in the context of SN progenitors, the high mass-loss rate of \kee 
faces difficulties in reproducing their position on the HRD due to the extreme stripping occurring even within a brief RSG phase.   
Conversely, \yang can reproduce the observed SN progenitor's positions. Note that a high mass-loss rate can be the cause of the high rate of observed events with low ejecta masses \citep{Martinez+2022}. In addition, due to the stripping of luminous RSGs, \yang predicts luminous YSG progenitors that are not observed in nearby SNe (Sect.~\ref{sec:SNprogenitors}). On the other hand, \dejager and \beasor are consistent with these detections.

Consequently, we find no single prescription that can reproduce all the observational constraints that we considered at the same time. We speculate that a RSG prescription with low on-average mass-loss rates but with a fast increase of the mass-loss rates towards high luminosities, 
may be needed to reproduce both the RSG luminosity function and the detected Type II SN progenitors.

In the analysis above, we acknowledge the presence of various uncertain physical effects and assumptions in our work, as well as in prior studies, which 
ought to be kept in mind and whenever possible carefully investigated.
The most significant of these factors are discussed below.

In the process of empirically inferring the mass-loss rates from fitting the infrared excess of the RSGs SEDs, various assumptions are made (which are not applicable for \kee that is theoretically derived). \yang assumed radiatively driven winds in DUSTY, where dust acts as the accelerating mechanism of mass loss. This assumption can lead to deriving higher mass-loss rates by two to three orders of magnitude than assuming a steady wind with a constant outflow velocity, where dust is considered a byproduct of the RSG mass loss \citep{Antoniadis+2024}. \citet{Beasor+2020} used steady-state winds which is one of the reasons that the rates are not as high as \yang.  The variation in grain sizes in the dust shell models could also lead to different rates by a factor of up to 20-30 \citep{Antoniadis+2024}. 

In addition, when modeling a dust shell, spherical symmetry is usually assumed. However, dust around RSGs is clumped and asymmetric \citep{Smith+2001, Scicluna+2015}. Luminosity measurements could be affected by dense shells, resulting in higher or lower observed luminosities depending on clump orientation. Furthermore, photometric variability in RSGs can induce luminosity variations of up to 0.2 dex \citep{Beasor+2021} due to obscuration \citep{Beasor+2022}. These effects influence the dusty RSGs at the high luminosity end, affecting the tail of the luminosity function and ratio of luminous RSGs.

At the same time, the dust occurrence and its effect in the stellar SED probes only the recent mass-loss rate of RSGs, which could be considered instantaneous relative to the overall evolution of the star. In the case of abrupt and episodic mass loss phenomena \citep{Montarges+2021, 
Munoz-Sanchez+2024}, the inferred rates would not be valid for the entire RSG lifetime, and implementing them as an average over the RSG lifetime overestimates the total mass lost. %

The mass-loss rate prescriptions are implemented in \posydon stellar models of \mesa, so the results are affected by the assumptions in the physical processes during their evolution \citep[explained in detail in][]{Fragos+2023, Andrews+inprep}. Notably, the single-star models used are assumed to be non-rotating, avoiding any extra rotationally-induced mixing throughout the evolution. At the same time, exponential convective core overshooting has been assumed, calibrated according to \citet{Brott+2011a}, which is tailored specifically for massive stars.
This assumed overshooting is greater than that incorporated in other 
stellar evolution tracks for RSGs, e.g. GENEC \citep{Ekstrom+2012} and MIST \citep{Choi+2016}, leading to high helium core masses and thus RSG luminosities \citep[e.g.,][]{Farrell+2020b}, and consequently in high average mass-loss rates for the same initial mass. 
At the same time, the surface temperatures of the convective envelopes of RSGs are governed by their evolution on the Hayashi tracks (unless they experience a blue loop to a radiative envelope or stripping of the envelope to go to the blue). The exact RSG structure for a given luminosity is dependent on various physical assumptions, most importantly the treatment of convective instability through mixing length theory \citep{Meynet+2015}. Thus, the effective temperatures of RSGs in our results should be sensitive to the adopted mixing length parameter, $\alpha_{\rm MLT} =1.93$ \citep{Fragos+2023}.


 It is important to note that the observed SN progenitor sample includes stars of varying metallicities with generally higher metallicity than solely SMC-like. Additionally, the binary history of a significant fraction of Type II SN progenitors could affect their final position on the HRD  \citep{Podsiadlowski+1992,  Zapartas+2019, Menon+2019, Menon+2024, Schneider+2024}.  Furthermore, possible obscuration of the SN progenitor due to pre-SN episodic mass loss \citep{Arnett+2011, Fuller2017} almost synchronized with the SN explosion may also contribute to the observed discrepancies  \citep{Davies+2022}. Evidence of such mechanisms has been observed, for example in the recent case of SN2023ixf \citep{Jacobson-Galan+2023, Kilpatrick+2023, Bostroem+2023}. 
The termination of our models at core carbon depletion, with less than a few or tens of years until core collapse,  does not account for any potential episodic mass loss that may occur during this remaining time. In any case, this is an uncertain mechanism with no well-constrained prescriptions regarding the probability of such events and the resulting mass-loss rate.

\subsection{Feedback}

\begin{figure}
\centering
\includegraphics[width=0.9\linewidth]{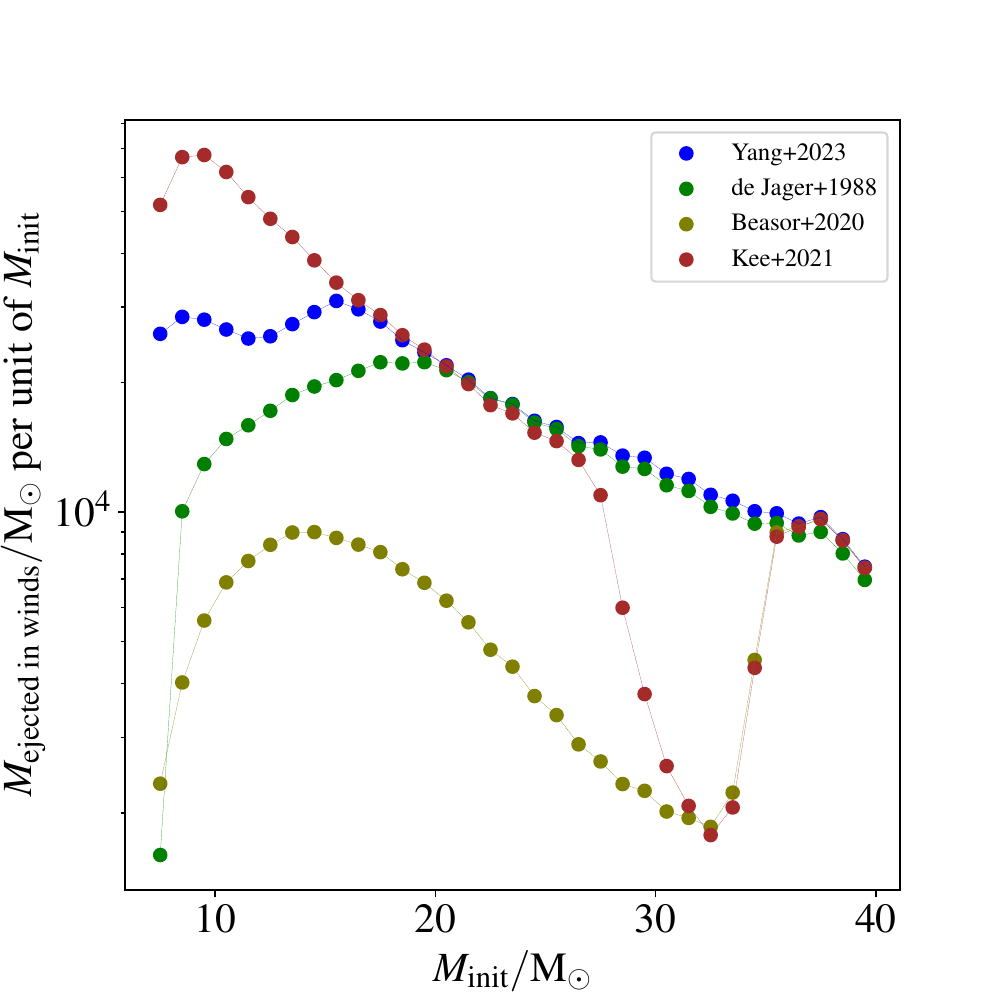}
\caption{Total mass lost during the stellar pre-SN lifetime, per  unit  initial mass bin, weighted with the IMF and following the SMC SFH.}
\label{fig:mass_loss_IMF_weighted}
\end{figure}

Massive stars play a crucial role in the evolution of galaxies by providing mechanical 
feedback and chemically enriching the interstellar medium \citep[e.g.,][]{Barkana+2001}. 
For stars below $M_{\rm init} \sim 25 \Msun$ the ejected mass before the explosion is dominated by the RSG phase (\fig{fig:Mlost_Time_Minit_andHRD} and \ref{fig:Mlost_Time_Minit_andHRD_appendix}).  
In \fig{fig:mass_loss_IMF_weighted} we calculate the total mass ejected from a stellar population as described in Sect.~\ref{sec:simulation_SMC}, i.e. weighted by the IMF and following the SFH of SMC. 
Different RSG wind prescriptions modify the feedback in several ways, influencing not only the total mass ejected during their pre-SN evolution, but also the initial masses that contribute most significantly and thus the timing of the feedback too as it depends on the evolutionary timescales of these masses. 
For example, following the high mass-loss rate prescriptions of \yang and \kee, the most important contributors in the SMC are the stars of $M_{\rm init} \sim 8-15 \Msun$ range, as these are favoured by the IMF. However, when implementing weaker wind prescriptions such as \beasor and \dejager, the dominant contribution shifts to stars with initial masses of $\sim 15-20 \Msun$ models. This shift occurs because lower-mass stars have smaller fractional mass loss according to these prescriptions (\fig{fig:Mlost_Time_Minit_andHRD} and \ref{fig:Mlost_Time_Minit_andHRD_appendix}), and more massive are rare  due to the IMF. 

The SN explosions are another important source of feedback,  which can drive outflows on galactic scales  \citep[e.g.][]{Shapiro+1976,Hopkins+2011}, 
and may trigger further star formation \citep{Krumholz+McKee2005}. \kee but also \yang predict smaller final hydrogen-rich envelope masses, diminishing the amount of mass expected to be ejected during the SN itself. This implies that, from a feedback perspective, the question of which stars will explode becomes less critical when we assume stronger RSG wind prescriptions, as most of the mass is ejected before the SN occurs.

\section{Summary and conclusions}
\label{sec:conclusions}

In this work, we investigate the impact of different red supergiant (RSG) mass-loss on the evolutionary pathways and observable properties of massive stars in the Small Magellanic Cloud (SMC), and their eventual roles as supernova (SN) progenitors. We achieved this by implementing in grids of single-stellar tracks within the \posydon framework, various RSG mass-loss prescriptions, inferred with different techniques and samples, and resulting in a wide range of mass-loss rates.   

We find that RSG mass loss prescriptions significantly affect the mass lost and the duration of the RSG phase. Higher mass-loss rates, such as in \citet{Kee+2021} or even \citet{Yang+2023} for the luminous ones, result in earlier envelope stripping and reduced RSG lifetimes, whereas lower rates, such as in  \citet{Beasor+2020}, prolong the RSG phase. The reason is that higher mass-loss rates cause RSGs to transition to a hotter phase at lower initial masses, down to $15 \msun$ for \citet{Yang+2023}. These stripped stars also avoid becoming progenitors of Type II SNe, influencing their final position in the Hertzsprung-Russell diagram and determining whether they will explode or implode directly as a black hole.

 Our findings indicate that none of the considered mass-loss prescriptions is fully consistent with all observational constraints simultaneously, highlighting the empirical complexities and theoretical uncertainties inherent in modeling RSGs. All the incorporated prescriptions have difficulty reconciling the observed distribution of RSG luminosities across the entire spectrum, particularly at lower luminosities, although this may be sensitive to observational biases. The high mass-loss rates suggested by \citet{Kee+2021} result in intensive stripping of the stellar envelope, leading to a predicted dearth of luminous RSGs and an excess of yellow supergiant SN progenitors, which are not commonly observed as progenitors of nearby supernovae. Conversely, the \citet{de-Jager+1988} and \citet{Beasor+2020} prescriptions slightly overestimate the number of luminous RSGs. Given a crude estimate of the effect of binarity, the above discrepancy would increase due to even more luminous binary products that will become RSGs. The empirical prescription of \citet{Yang+2023} seems more consistent with the observed luminosity function, naturally reproducing the updated Humphreys-Davidson limit of $\logL\sim 5.5$ for the Magellanic Clouds \citep{Davies+2018b} and M31 \citep{McDonald+2022}. This highlights the possible importance of a turning point to increased mass-loss rates for luminous RSGs, as suggested also by \citet{Vink+2023} and \citet{Antoniadis+2024}. In contrast, the stripping of the highly luminous RSGs according to \citet{Yang+2023}, which predicts yellow supergiant SN progenitors, seems inconsistent with detected ones.

In conclusion, our study highlights the importance of constraining RSG mass loss with more complete observational samples especially at lower luminosities, and a clear investigation of the model uncertainties \citep[e.g.,][]{Antoniadis+2024}, the dependence on other parameters such as the gas-to-dust ratio, the outflow speed \citep[e.g.,][]{Goldman+2017} or the stellar mass \citep{Beasor+2020}. 
Further observational constraints \citep[e.g.,][]{Decin+2024} are crucial for refining the important theoretical attempts to model RSG mass loss \citep{Kee+2021,Vink+2023,Fuller+2024} and eventually understanding the causing mechanism. Finally, studies of the effect of possible binary-induced or episodic mass loss \citep{Cheng+2024}, during the RSG phase or before the SN, would enhance our insight into the late stages of massive stellar evolution.

\begin{acknowledgements}

The authors thank Ming Yang for helpful discussions and for providing the data of the 2023 work he led, prior to publication.  They also thank Dylan Kee, Alex de Koter and Evangelia Christodoulou for useful discussions. 
EZ and DS acknowledge support from the Hellenic Foundation for Research and Innovation (H.F.R.I.) under the “3rd Call for H.F.R.I. Research Projects to support Post-Doctoral Researchers” (Project No: 7933). EZ, SdW, KA, GMS, AB, GM acknowledge funding support from the European Research Council (ERC) under the European Union’s Horizon 2020 research and innovation programme (``ASSESS", Grant agreement No. 772086). 
The POSYDON project is supported primarily by two sources: the Swiss National Science Foundation (PI Fragos, project numbers PP00P2\_211006 and CRSII5\_213497) and the Gordon and Betty Moore Foundation (PI Kalogera, grant award GBMF8477). MB acknowledges support from the Boninchi Foundation. 
KAR is also supported by the Riedel Family Fellowship and thanks the LSSTC Data Science Fellowship Program, which is funded by LSSTC, NSF Cybertraining Grant No.\ 1829740, the Brinson Foundation, and the Moore Foundation; their participation in the program has benefited this work. 
KK is supported by a fellowship program at the Institute of Space Sciences (ICE-CSIC) funded by the program Unidad de Excelencia María de Maeztu CEX2020-001058-M. 
JJA acknowledges support for Program number (JWST-AR-04369.001-A) provided through a grant from the STScI under NASA contract NAS5-03127. ZX acknowledges support from the China Scholarship Council (CSC).

\end{acknowledgements}

\bibliographystyle{aa}
\bibliography{my_bib,bib_overleaf}

\begin{appendix}

\section{Average Red Supergiant luminosity as a function of initial mass}\label{sec:Minit_LRSG}

\begin{figure}
\centering
\includegraphics[width=1.15\linewidth]{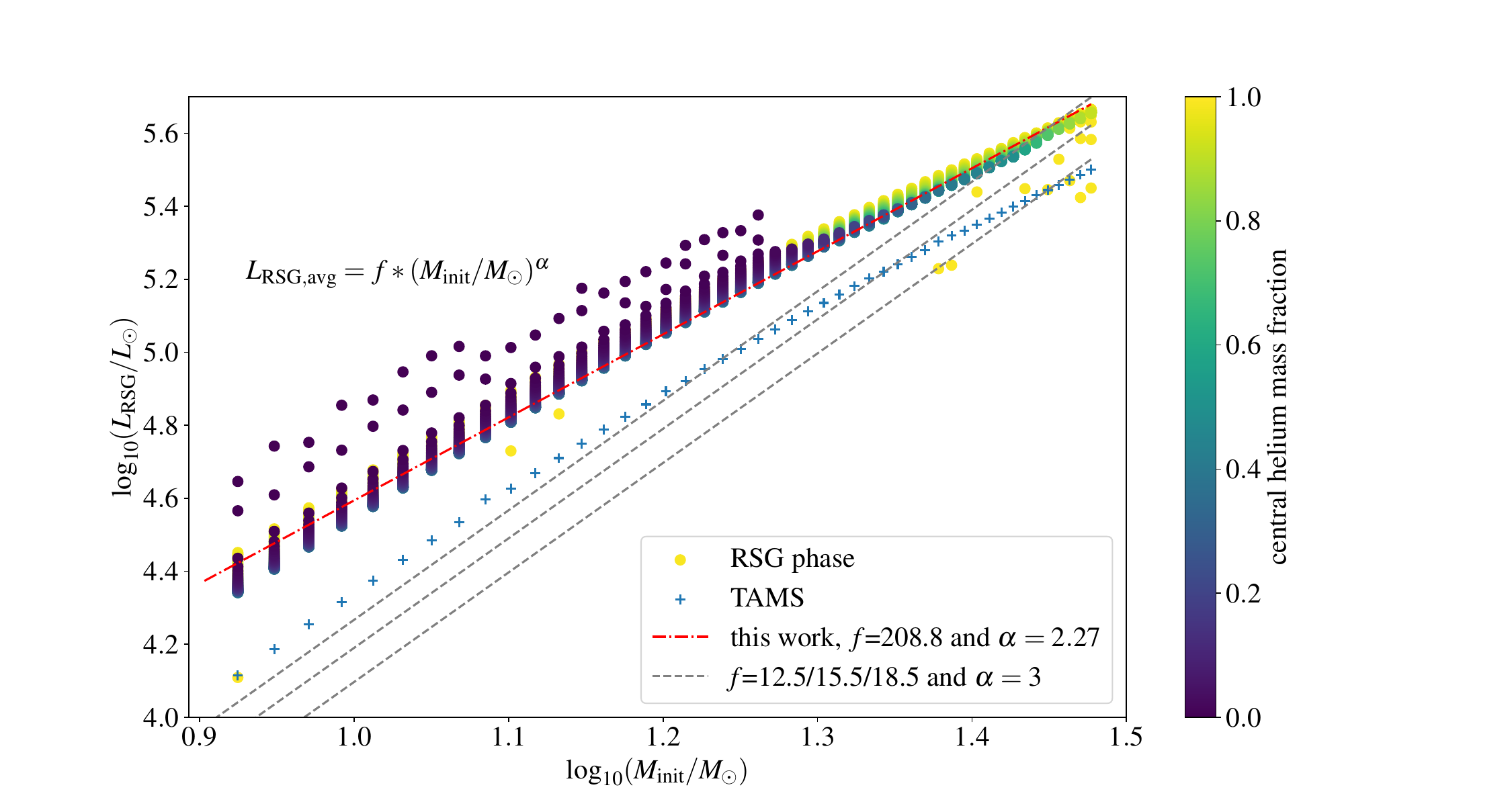}
\caption{Luminosity during the RSG phase as a function of the initial mass of the star. The points are distributed in 100 equal timesteps that span the whole time range of each model during its RSG phase, colored by the central helium mass fraction. For reference, we also show the luminosity at TAMS and various fittings from this or previous works, of the form of Eq.~\ref{eq:Minit_L}.
}
\label{fig:Minit_LRSG}
\end{figure}

In \fig{fig:Minit_LRSG} we show the luminosity of stars during their post-MS evolution. 
Each streak of vertical dots represents the luminosity during the RSG phase for a given $M_{\rm init}$ in 100 points equally distributed in time. We also show the luminosity of that model at the end of MS (TAMS), just before becoming a RSG. 
The luminosity shows an increase from TAMS to around $\gtrsim 0.5$ dex when it reaches the RSG phase and then quickly increases a bit. The star keeps an almost constant luminosity during the longer timescale of core helium burning, increasing again only after helium depletion (for initially lower mass stars) or after being stripped and ending their RSG phase (for initially higher mass stars). During that phase, the RSG may change its temperature (even doing a blue loop), and thus part of its evolution with $T_{\rm eff} > T{\rm min}$ will not be included. 

We fit the time-averaged $L_{\rm RSG,avg}$ as function of $M_{\rm init}$, using the same equation as in \citet{Kee+2021}:

\begin{equation}\label{eq:Minit_L}
L_{\rm RSG,avg}=f\cdot(M_{\rm init}/M_{\odot})^{\alpha}
\end{equation}

The relation they use in that work has $\alpha = 3$, with the factor $f$ ranging from 12.5 for the GENEVA models \citep{Ekstrom+2012} to 18.5 for the default MESA models \citep{Paxton+2011}, eventually picking an average value.  We show these relations in \fig{fig:Minit_LRSG}.  In this study, following \dejager, we fit the on average higher luminosity values during the RSG phase, resulting in a shallower slope of $2.27$ and a much higher $f=208.6$.

Interestingly, we found the relation not to be highly dependent on the RSG mass loss prescription applied, especially the slope  $\alpha$ which takes the value of  2.25, 2.26, 2.18, and $f$ factor becoming 225, 214.9, 294.1  for the \yang, \beasor, \kee, respectively. This is probably caused by the fact that luminosities are more tightly correlated to the helium core mass of the star, which is formed at TAMS and relatively unaffected by the amount of mass of the surrounding envelope \citep{Justham+2014,Farrell+2020b}. Of course, it is still sensitive to the model's uncertain physical assumptions, including convective overshooting and rotational mixing.

We confirm the loose correlation of the initial or current total mass of a RSG with its luminosity, but here we want to correspond a time-averaged luminosity during the RSG phase for a given initial mass. However, as discussed in Sect.~\ref{sec:mass_lost_time} the time spent during the phase is highly influenced by the wind mass loss prescription.

\section{Temperature and Mass loss of RSGs}

\begin{figure*}
\centering
\includegraphics[width=0.5\linewidth]{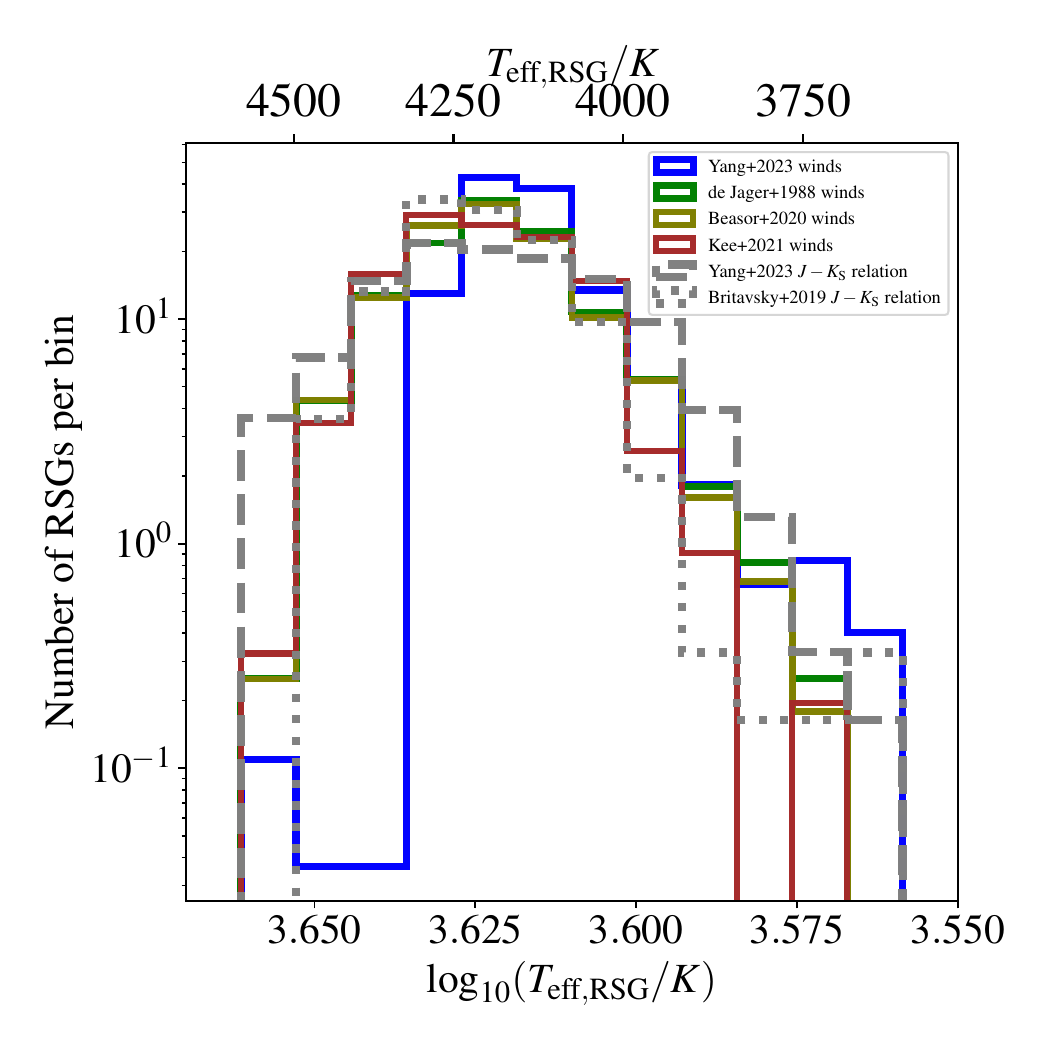}\includegraphics[width=0.5\linewidth]{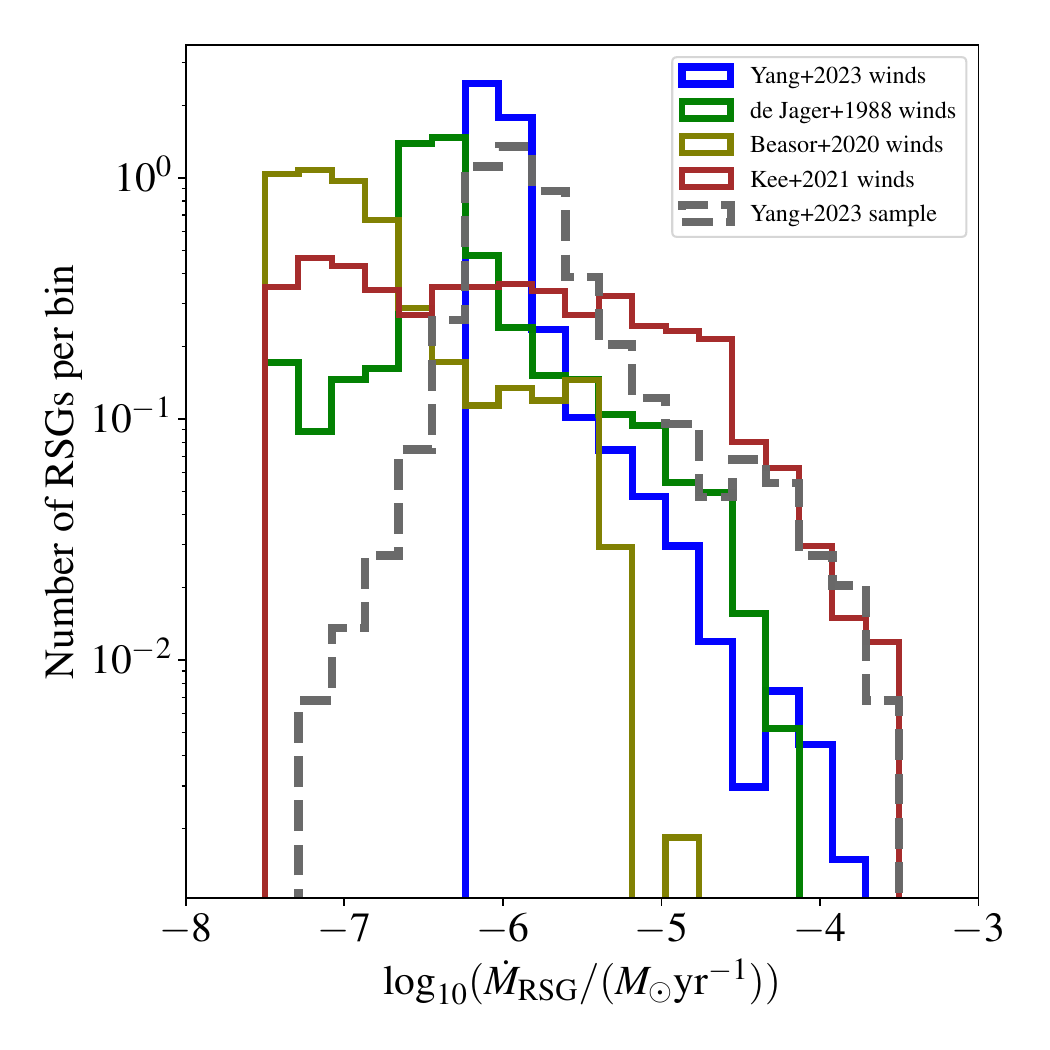}
\caption{Normalized distributions of a) effective temperatures b) mass-loss rates of the \posydon populations of RSGs, for all the different mass loss prescriptions, compared against the observationally inferred values.}
\label{fig:Teff_Mdot_distr}
\end{figure*}

In \fig{fig:Teff_Mdot_distr} we show the effective temperature and the mass-loss rate of RSGs for the different prescriptions, compared with the RSG sample of \yangt, with the limits as discussed in Sect.~\ref{sec:obs_sample}. 
$T_{\rm eff, RSG}$ of the observed sample is inferred from $J-K_{\rm S}$ color of the sample, using relation from \citet{Yang+2020} and \citet{Britavskiy+2019}.

It  is also interesting that models following \yang have  a more peaked mass-loss rate 
 distribution than the observationally inferred one of \yangt, where this prescription was based. This is mainly because the prescription is only luminosity dependent and thus cannot fully capture the spread of the inferred mass-loss rates of the sample for a given luminosity (as depicted in \fig{fig:Mdot_luminosity}). 

\section{Relation between initial and Terminal-Age Main-Sequence mass}\label{sec:TAMS}

\begin{figure}
\centering
\includegraphics[width=\linewidth]{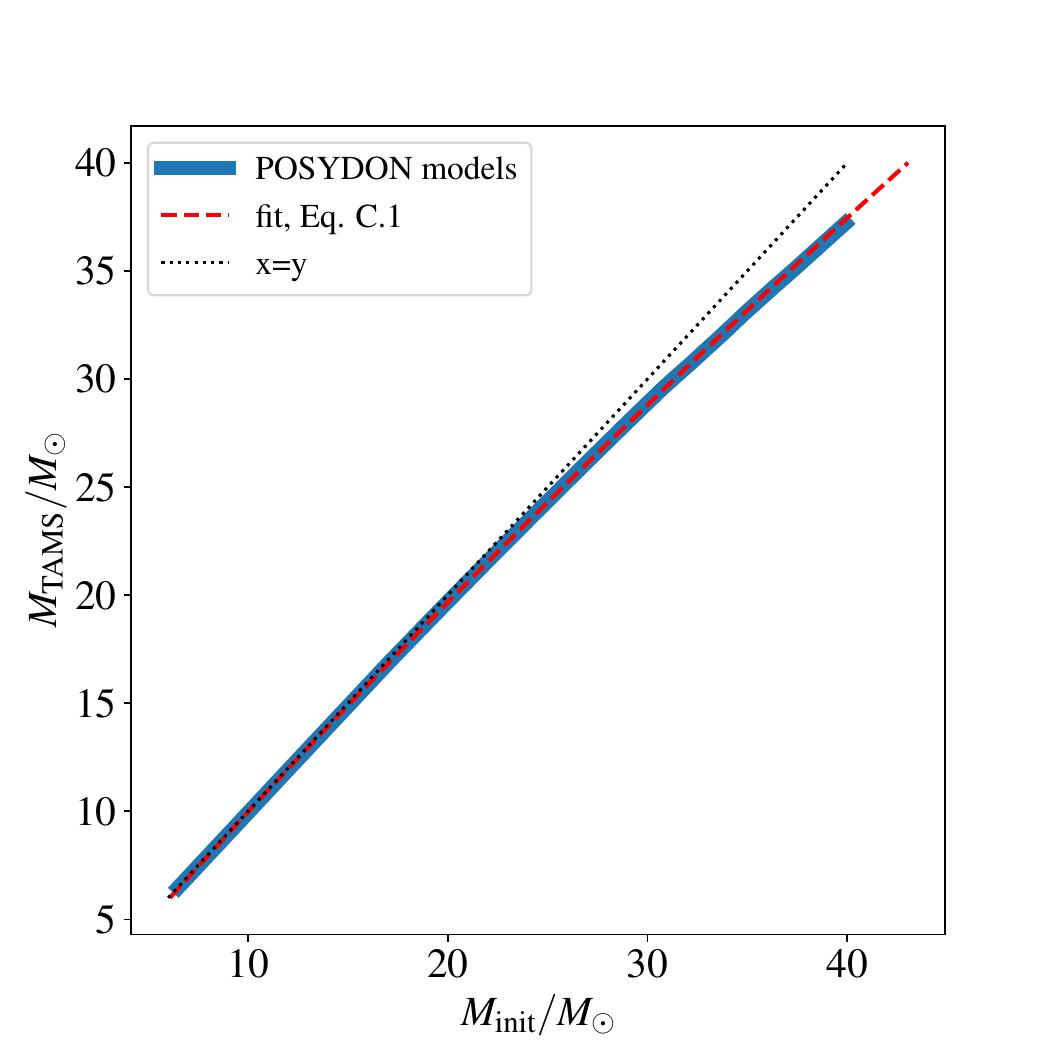}
\caption{Relation between initial and TAMS mass for our \posydon models.}
\label{fig:TAMS_ZAMS_mass}
\end{figure}

In \fig{fig:TAMS_ZAMS_mass} we present the relation between initial and TAMS mass for our \posydon models which was used in Sect.~\ref{sec:binarity} to take into account binary products in the luminosity function of RSGs. We choose $M_{\rm TAMS}$ as a proxy of the core mass and luminosity during the RSG phase for binary products, and we extra the following  fitting relation with $M_{\rm init}$ :

\begin{equation}\label{eq:TAMS_ZAMS_mass}
M_{\rm init}/\Msun = 
0.391 +  0.923\cdot(M_{\rm TAMS}/\Msun) + 0.0034\cdot(M_{\rm TAMS}/\Msun)^2 
\end{equation}

\end{appendix}

\end{document}